\documentclass[10pt,final,doublecolumn]{IEEEtran}
\usepackage{amsmath}
\usepackage{amssymb}
\usepackage{amsfonts}
\usepackage{latexsym}
\usepackage{graphicx}
\usepackage{bbding}
\usepackage{enumerate}
\usepackage{subeqnarray}
\usepackage{multicol}
\usepackage{color}
\usepackage{slashbox}
\usepackage{setspace}
\usepackage{mathrsfs}
\usepackage{array}
\usepackage{algorithm,algorithmic}
\usepackage{colortbl}
\usepackage{bm}
\IEEEoverridecommandlockouts
\allowdisplaybreaks[4]

\begin{document}
\title{Robust Design of Federated Learning for Edge-Intelligent Networks}
\author{Qiao Qi and Xiaoming Chen
\thanks{Qiao Qi ({\tt qiqiao1996@zju.edu.cn}) and Xiaoming Chen ({\tt chen\_xiaoming@zju.edu.cn}) are with the College of Information Science and Electronic Engineering, Zhejiang University, Hangzhou 310027, China.}}\maketitle

\begin{abstract}
Mass data traffics, low-latency wireless services and advanced artificial intelligence (AI) technologies have driven the emergence of a new paradigm for wireless networks, namely edge-intelligent networks, which are more efficient and flexible than traditional cloud-intelligent networks. Considering users' privacy, model sharing-based federated learning (FL) that migrates model parameters but not private data from edge devices to a central cloud is particularly attractive for edge-intelligent networks. Due to multiple rounds of iterative updating of high-dimensional model parameters between base station (BS) and edge devices, the communication reliability is a critical issue of FL for edge-intelligent networks. We reveal the impacts of the errors generated during model broadcast and model aggregation via wireless channels caused by channel fading, interference and noise on the accuracy of FL, especially when there exists channel uncertainty. To alleviate the impacts, we propose a robust FL algorithm for edge-intelligent networks with channel uncertainty, which is formulated as a worst-case optimization problem with joint device selection and transceiver design. Finally, simulation results validate the robustness and effectiveness of the proposed algorithm.
\end{abstract}


\section{Introduction}

The sixth-generation (6G) wireless networks are desired to support a variety of advanced wireless services, such as AR/VR, smart city, and video/audio surveillance, which generate a vast amount of data at edge devices \cite{6G1}-\cite{6G3}. In order to decrease transmission delay and communication burden, it is appealing to process these data at the network edge, but not the cloud servers. Driven by such a requirement, many researchers predict that 6G wireless networks will be edge-intelligent networks powered by the integration of edge computing and machine learning \cite{6G4}, \cite{6G5}.

Actually, edge intelligence has attracted considerable interests from both academia and industry. Traditional data-sharing-based edge-intelligent networks require to convey a large volume of data from edge devices to the base station (BS) for training a global model \cite{CL1}, \cite{CL2}. However, it not only leads to network congestion due to mass data transmission, but also causes the personal privacy leakage. With the rapid development of information technology and chip technology, the federated edge learning (FEEL) framework has been developed to address these issues, which implements distributed machine learning at the network edge (e.g., intelligent devices and the BS) \cite{FEEL1}. In this context, model-sharing-based federated learning (FL) with the nature of privacy protection has emerged as a new branch for edge-intelligent networks \cite{FL1}. Different from conventional edge intelligence supported by data sharing, FL exploits the computation capabilities of multiple intelligent devices, which are coordinated by an edge server, e.g., the BS, to cooperatively train a global model in an iterative manner \cite{FL2}.  Specifically, in each iteration (also called a communication round) of FL, the edge server first broadcasts the global model parameters to the devices, such that all participating devices can synchronize and initialize their local models. After that, each participating device individually trains and updates the local model based on its own dataset, and then sends the updated local model parameters to the edge server for global model aggregation. Since the whole training process of the global model only involves the transmission of model parameters rather than real raw data, FL is very appealing to edge-intelligent networks for data analysis and processing with better privacy.

\subsection{Related Works and Motivation}

Since the exchange of model parameters between the server and the devices is a key process of FL, the design of communication-efficient FL has received wide attention in recent years. As a leading and widely used algorithm in FL, federated averaging (FedAvg), also called model-average FL, runs stochastic gradient descent (SGD) in parallel on a portion of the total devices and obtains the global model by averaging the model parameters, which effectively reduces the number of communication rounds for global model aggregation \cite{FL3}.  In particular, FedAvg has been shown to work in real-world settings by Google with their GBoard in next-word-prediction \cite{GBoard1} and emjoi-prediction \cite{GBoard2}.  Based on FedAvg, many excellent FL algorithms have emerged.  For example, the federated proximal (FedProx) algorithm was proposed by adding regularization on each local loss function \cite{FL4}. Additionally,  Acar et. al. introduced the federated learning with dynamic regularization (FedDyn) algorithm as a solution to heterogenous dataset setting \cite{FL5}, which dynamically regularizes each devices' loss function such that the modified device losses converge to the actual global loss.

In practice, edge devices are usually connected through long-distance wireless access. Due to limited radio resources and iterative transmissions of high-dimensional model parameters, especially when applying deep neural networks (DNNs) into model training as DNNs nowadays often have millions of parameters \cite{ML0}, the high communication burden is a main challenge of FL for edge-intelligent networks. To this end, some prior works attempted to solve it in various aspects. From the perspective of machine learning, it is usual to reduce the communication load by decreasing the number of participating devices or compressing the model parameters via quantization and sparsification. For example, the authors in \cite{ML1} proposed a joint learning and communication framework for FL by resource allocation and device selection with the goal of minimizing an FL loss function. In \cite{ML2}, an universal vector quantization scheme for FL was put forward to effectively reduce the communication-latency. These methods from the aspect of machine learning regard wireless channels as a set of independent error-free bit-limited links between the devices and the server. In fact, the transmission of model parameters over wireless channels are inevitably affected by channel fading, interference, and noise. As a result, the performance of FL is unsatisfactory. Thus, it is necessary to develop novel transmission schemes to achieve communication-efficient FL.

During the transmission of model parameters in FL, co-channel interference has a great impact. To avoid the interference and achieve accurate model aggregation, a common way is to adopt an orthogonal multiple access (OMA) scheme, i.e., allocating an orthogonal resource block to each participating device for uploading its local model parameters \cite{OMA1}-\cite{OMA3}. In \cite{OMA1}, the authors studied a probabilistic scheduling framework of FL based on orthogonal frequency-division multiple access (OFDMA). Besides, energy-efficient radio resource allocation algorithms for FL over wireless fading channels supported by OMA schemes were proposed in \cite{OMA2} and \cite{OMA3}. Yet, such approaches substantially restrict the number of accessible devices that participate in FL under limited spectrum resources. In particular, it causes a high transmission latency and a low spectrum efficiency when the number of participating devices for FL is large, and thus is no longer applicable for edge-intelligent networks with a massive number of devices. In this context, a promising solution called \emph{over-the-air computation} (AirComp) has been proposed and raised wide interests \cite{AirComp1, AirComp2}, which exploits the superposition property of wireless multiple access channels (MACs) to compute a class of \emph{nomographic functions} of distributed data from devices via concurrent transmission \cite{Nomofun}. Fortunately, the averaging method for global model aggregation in the FedAvg algorithm falls into the category of nomographic functions.  Hence, it is possible to effectively enhance the learning performance by applying AirComp into model-average FL for fast model aggregation and accurate learning performance. For example, the authors in \cite{Heterogeneous} developed a convergent over-the-air FL algorithm by introducing a time-varying precoding and scaling scheme. In \cite{Onebit}, a novel digital over-the-air aggregation scheme was proposed to alleviate the effects of wireless channel hostilities on the convergence rate of FL.

To achieve accurate model aggregation based on AirComp over fading channels, the devices need to have channel state information (CSI) to carry out channel pre-equalization. Yet, inaccurate channel estimation and hardware impairment at the devices may cause imperfect equalization, and thereby distort the aggregated model. Some works have taken into consideration the aggregation error via uplink and designed efficient schemes to improve the system performance.  For example, the authors in \cite{BBAFL} derived the trade-off between communication and learning, and proposed a broadband analog aggregation scheme for FL to reduce the communication-latency. In \cite{YKFL}, a resource allocation scheme for AirComp-based on-device distributed FL system was proposed to reduce the aggregation error and improve the learning performance when the edge server was equipped with multiple antennas. In \cite{IRSFL}, the authors proposed a novel AirComp-based FL system to achieve fast and reliable model aggregation by utilizing the intelligent reflecting surface to compensate for the magnitude reduction and misalignment of AirComp. However, most of exiting works for Aircomp-based FL mainly focus on the effect of parameter transmission via uplink while overlooking the impact of downlink parameters transmission for model broadcast. In fact, due to imperfect channel estimation, feedback quantization, or delay in signal acquisition on fading channels, the received model parameters at the devices may not be perfectly recovered. Such model distortion caused by unfavourable wireless transmission leads to performance degradation in terms of convergence speed and prediction accuracy. This key observation is also noted in a recent parallel work \cite{parallel}, which assumed an error-free parameter aggregation over uplink and focused on studying the impact of noisy downlink on the FL performance. Then, an analog downlink transmission approach was proposed and the convergence of FL for the proposed downlink approach was well studied.  In order to compare related works more clearly, we list a comparison table to highlight the contribution of our work, cf. Table \ref{comp}. Besides, the number of participating devices at each communication round is also a critical factor of the FL performance, which has different impacts on learning and communication. On the one hand, selecting more devices to participate in FL can collect more information from distributed devices by model aggregation, which is benefit to the convergence of global model. On the other hand, the error caused by AirComp-based model aggregation via wireless transmission increases as the number of participating devices for FL increases, which in turn results in the degradation of learning performance. Thus, it is necessary to characterize the effects of the number of participating devices and the distortion of transmitted model on the performance of FL, and then design a corresponding transmission scheme for FL in edge-intelligent networks.

\begin{table*}[t]
\small
\centering
\caption{The comparison of related works}
\label{comp}
\begin{tabular}{|m{0.17\textwidth}<{\centering}|m{0.23\textwidth}<{\centering}|m{0.27\textwidth}<{\centering}|m{0.24\textwidth}<{\centering}|}%
    \hline
        ~ & \cite{parallel} & \cite{YKFL}& Our work \\ \hline
        FL Model & Gradient-average FL & Model-average FL & Model-average FL\\ \hline
        Uplink Error & No & Yes & Yes  \\ \hline
        Downlink Error & Yes & No & Yes  \\ \hline
        CSI & Perfect & Perfect & Imperfect  \\ \hline
        Transceiver & Downlink: zero-forcing receiver at the device & Uplink: receiver at the BS based on the optimization problem with a given zero-forcing transmitter at the device & Uplink/Downlink: adaptive transceivers based on the optimization problem  \\ \hline
        Objective & Analyzing the convergence of  FL for the proposed downlink transmission approach & Improving the FL performance by optimizing device selection and the receiver at the BS & Improving the FL performance by optimizing device selection and transceiver design  \\ \hline
        \end{tabular}
\end{table*}

%\begin{table}[t]
%\centering
%\caption{The comparison of related works}
%\label{comp}
%\begin{tabular}{|m{0.17\textwidth}<{\centering}|m{0.23\textwidth}<{\centering}|m{0.27\textwidth}<{\centering}|m{0.24\textwidth}<{\centering}|}%
%    \hline
%        ~ & \cite{parallel} & \cite{YKFL}& Our work  \\ \hline
%       FL Model & Gradient-average FL & Model-average FL & Model-average FL\\ \hline
%        Uplink Error & No & Yes & Yes  \\ \hline
%        Downlink Error & Yes & No & Yes  \\ \hline
%        CSI & Perfect & Perfect & Imperfect  \\ \hline
%        Transceiver & Downlink: zero-forcing receiver at the device & Uplink: receiver at the BS based on the optimization problem with a given zero-forcing %transmitter at the device & Uplink/Downlink: adaptive transceivers based on the optimization problem  \\ \hline
%        Objective & Analyzing the convergence of  FL for the proposed downlink transmission approach & Improving the FL performance by optimizing device selection and the receiver at the BS & Improving the FL performance by optimizing device selection and transceiver design  \\ \hline
%        \end{tabular}
%\end{table}

\subsection{Contributions}
In this paper, we consider a practical edge-intelligent network with imperfect CSI, inevitably leading to errors of both model aggregation and model broadcast in the training process of FL. To improve the performance of model-average FL via AirComp, we provide a robust transmission scheme for FL in edge-intelligent networks. The contributions of this paper are three-fold:

\begin{enumerate}

\item We propose a communication-efficient FL framework for edge-intelligent networks under practical but adverse conditions, where the BS as an edge sever only has partial CSI due to imperfect channel estimation or feedback. To realize fast model aggregation, AirComp technique is applied into model-average FL for effectively improving the communication efficiency over limited radio resources.

\item We study the distortion in the received models at both the BS and the devices caused by wireless transmission and reveal the impacts of the errors of model parameter transmission and the number of participating devices on the performance of FL. Specifically, the combination of the broadcast error over the downlink and the aggregation error over the uplink may lead to a notable degradation of learning performance, while the convergence of training can be accelerated with more involved devices.

\item We provide a robust transmission scheme with joint device selection and transceiver design to improve the performance of FL. This design is formulated as a computationally difficult combinatorial optimization problem, which maximizes the number of selected devices while satisfying the MSE requirements of model broadcast via downlink and model aggregation via uplink, respectively. Then, an effective algorithm is proposed to solve such a complicated nonconvex problem and obtain a feasible sub-optimal solution.
\end{enumerate}

\subsection{Organization and Notations}
The rest of this paper is organized as follows: Section II gives a concise introduction of a practical FL framework for  edge-intelligent networks. Section III focuses on the robust design of a communication-efficient transmission scheme for FL. Section IV presents extensive simulation results to validate the effectiveness and robustness of the proposed scheme. Finally, Section V summarizes the paper.

\emph{Notations}: We use bold upper (lower) letters to denote matrices (column vectors), $(\cdot)^H$ to denote conjugate transpose, $\|\cdot\|_0$ to denote the $\ell_0$-norm of a vector, $\|\cdot\|_1$ to denote the $\ell_1$-norm of a vector, $\|\cdot\|$ to denote the $\ell_2$-norm of a vector, $|\cdot|$ to denote the absolute value of a scalar or the size of a set, $\mathrm{Re}\{\cdot\}$ to denote the real parts of matrices,  ${{\mathbb{C}}^{M\times N}}$ to denote the set of $M$-by-$N$ dimensional complex matrix, ${{\mathbb{R}}^{m\times n}}$ to denote the set of $m$-by-$n$ dimensional real matrix,  $\mathbb{E}\{\cdot\}$ to denote expectation, and $\mathcal{CN}(\mu,\sigma^2)$ to denote the circularly symmetric complex Gaussian (CSCG) distribution with mean $\mu$ and variance $\sigma^2$.

\section{System Model}
\begin{figure}[h] \centering
\includegraphics [width=0.5\textwidth] {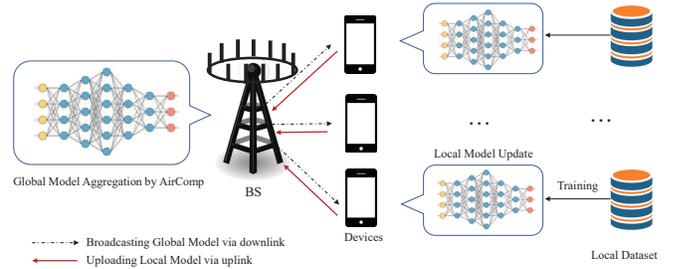}
\caption {A model of an edge-intelligent network.}
\label{Fig1}
\end{figure}

Consider a single-cell edge-intelligent network operated in the time division duplex (TDD) mode\footnote{The proposed scheme can be directly extended to the multiple-cell scenario by taking into account inter-cell interference.}, cf. Fig. \ref{Fig1}, which consists of a BS equipped with $N$ antennas serving as an edge server, and $K$ single-antenna intelligent devices. A computation model for data analysis and processing at the BS is trained with the aid of intelligent devices based on FL.

\subsection{FL Model}

For FL, a shared learning process with a global model $\mathbf{q}$ is trained cooperatively by multiple selected intelligent devices. Each selected device $k\in\mathcal{K}=\{1,2,\ldots,K\}$ has its own local dataset $\mathcal{D}_{k}$ with labeled data samples $\left\{\left(\mathbf{g}_{j}, l_{j}\right)\right\} \in \mathcal{D}_{k}$, where $\left(\mathbf{g}_{i}, l_{i}\right)$ denotes the input-output data pair consisting of training vector sample $\mathbf{g}_{i}$ and its ground-truth label $l_{i}$.  The sample-wise loss function is defined as $f_i(\mathbf{q};\mathbf{g}_i,l_i)$ for sample $i$. Then, the global loss function on all distributed datasets can be expressed as
\begin{equation}\label{eq1}
\mathcal{F}\left( \mathbf{q} \right)=\frac{1}{\left| {{D}_{\text{total}}} \right|}\sum\limits_{j\in {{D}_{\text{total}}}}{f_j\left( \mathbf{q};{{\mathbf{g}}_{j}},{{l}_{j}} \right)},
\end{equation}
where ${{D}_{\text{total}}}=\left\{ {{D}_{1}},{{D}_{2}},\cdots ,{{D}_{K}} \right\}$, and $\left| {{D}_{\text{total}}} \right|$ is the size of dataset ${{D}_{\text{total}}}$. The learning target is to optimize the model $\mathbf{q}$ by minimizing the global loss function $\mathcal{F}(\mathbf{q})$, i.e.,
\begin{equation}\label{eq2}
\mathbf{q}^{*}=\arg \min \mathcal{F}(\mathbf{q}).
\end{equation}
The model-averaging FL algorithm is commonly employed to find the optimal model $\mathbf{q}^{*}$ in an iterative manner, which obtains the global model $\mathbf{q}$ by averaging the local models $\mathbf{q}_{k}$ without the need of sharing datasets among the intelligent devices.
%As demonstrated in Fig. \ref{Fig1}, the FL algorithm alternates between two stages. In the first stage, the local models trained at selected devices are simultaneously sent to the BS for model-averaging via uplinks, then the BS updates the global model $\mathbf{q}$. In the second stage, the BS broadcasts the updated global model to all devices via downlink, then each selected device updates the local model by minimizing the local loss function with its own dataset. Specifically, the model updating rules as follows:
%\begin{subequations}\label{eq3}
%  \begin{eqnarray}
%    \text{BS: }&&\mathbf{q}=\frac{1}{|D_{\text{total}}|}\sum\limits_{k=1}^{K}{|{{D}_{k}|}{{\mathbf{q}}_{k}}},\\
%    \text{Devices: }&&{{\mathbf{q}}_{k}}=\arg \min {{F}_{k}}\left( \mathbf{q} \right),k=1,\cdots K,
%  \end{eqnarray}
%\end{subequations}
%where ${{\mathbf{q}}_{k}}$ is the local model at the $k$th device and ${{F}_{k}}\left( \mathbf{q} \right)$ is the local loss function for the $k$th device with dataset $D_k$, which can be expressed as
%\begin{equation}\label{eq4}
%  F_{k}(\mathbf{q})=\frac{1}{|D_{k}|} \sum_{\left(\mathbf{x}_{i}, y_{i}\right) \in \mathcal{D}_{k}} f_i\left(\mathbf{q} ; \mathbf{x}_{i}, y_{i}\right)
%\end{equation}
Due to the transmission of model parameters between the BS and the devices via wireless links, received model parameters may be distorted. For example, imperfect channel estimation, feedback quantization, or delay in signal acquisition on fading channels may lead to imperfect recovery of model parameters at the BS and at the devices. In general, the distortion of the received model at the devices and the BS in the $t$-th iteration can be modeled as
\begin{subequations}\label{eq6}
  \begin{eqnarray}
  \text{Devices: }&&\mathbf{q}_{i}^{[t]}=\mathbf{q}^{[t-1]}+\Delta\mathbf{q}_i^{[t]},i\in{\mathcal{S}_{t}},\\
   \text{BS: }&&\mathbf{q}^{[t]}=\frac{1}{\sum\limits_{i\in \mathcal{S}_t}{\left| {{D}_{i}} \right|}}\sum\limits_{i\in \mathcal{S}_t}{ {\left| {{D}_{i}} \right|}\mathbf{q}_i^{[t]}}+\Delta \mathbf{q}^{[t]},
  \end{eqnarray}
\end{subequations}
where $\mathcal{S}_t$ is the selected device set in the $t$-th iteration, $\Delta\mathbf{q}_i^{[t]}$ and $\Delta\mathbf{q}^{[t]}$ are the broadcast error at the $i$-th device and the aggregation error at the BS in the $t$-th iteration, respectively. In summary, the model-average FL algorithm is described as Algorithm \ref{algFL}. Note that for step 4 of Algorithm \ref{algFL}, a single-step SGD method updates the local model, which is given by
\begin{equation}\label{eqSGD}
\mathbf{q}_{i}^{[t]}=\mathbf{{q}}_{i}^{[t-1]}-\eta_i \nabla {\mathcal{F}_{i}}\left( \mathbf{{q}}_{i}^{[t-1]} \right),
\end{equation}
where $\eta_i$ is the learning step size and $\nabla {\mathcal{F}_{i}}\left( \mathbf{{q}}_{i}^{[t-1]} \right)$ is the gradient of ${\mathcal{F}_{i}}\left( \mathbf{{q}}_{i}^{[t-1]} \right)$ at the point $\mathbf{{q}}_{i}^{[t-1]} $. Then, the updated local model $\mathbf{q}_{i}^{[t]}$ can be obtained until convergence by $\tau$-step ($\tau\geq1$) SGD repeating the updating rule in (\ref{eqSGD}).

\begin{algorithm}[t]
\caption{: Model-average FL algorithm for edge-intelligent networks.}
\label{algFL}
\begin{algorithmic}[1]
\STATE{\textbf{Initialize} global model $\mathbf{q}^{[0]}$, communication round index $t=1$;}
\REPEAT
 \STATE{The BS determines the participating devices set $\mathcal{S}_t \in \mathcal{K}$;}
 \STATE{The BS broadcasts the current global model $\mathbf{q}^{[t-1]}$ to the participating devices via downlink.}
 \STATE{Based on the received model $\mathbf{q}_i^{[t]}=\mathbf{q}^{[t-1]}+\Delta\mathbf{q}_i^{[t]}$, each device $i \in \mathcal{S}_t$ performs a SGD algorithm to update the local model with its own dataset $D_i$.}
 \STATE{Participating devices upload their updated local models $\mathbf{q}_i^{[t]}$ to the BS via uplink; }
 \STATE{The BS updates the global model as $\mathbf{q}^{[t]}=\frac{1}{\sum\limits_{i\in \mathcal{S}_t}{\left| {{D}_{i}} \right|}}\sum\limits_{i\in \mathcal{S}_t}{ {\left| {{D}_{i}} \right|}\mathbf{q}_i^{[t]}}+\Delta \mathbf{q}^{[t]}$; }
 \STATE{$t=t+1$;}
\UNTIL{Global model convergence}
 \end{algorithmic}
\end{algorithm}

\begin{figure}[h]
\centering
\includegraphics [width=0.5\textwidth] {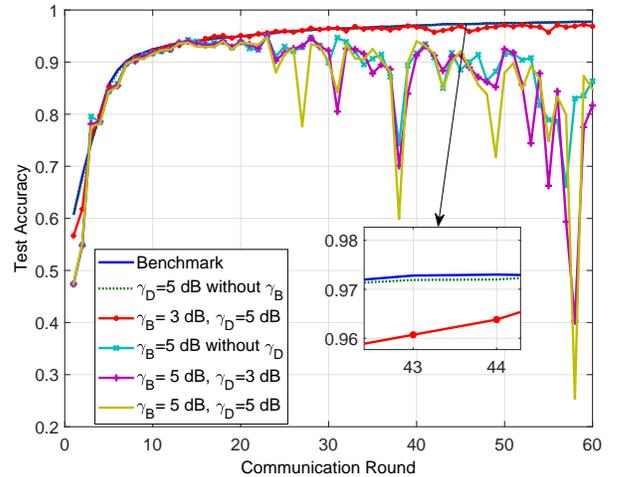}
\caption {The test accuracy of FL versus communication rounds under different error situation with all devices, where $\gamma_D$ and $\gamma_B$ are the MSE at the devices and at the BS.}
\label{Fig_MSE}
\end{figure}

In the following, we train a convolutional neural network (CNN) on the commonly used MNIST dataset with the model-average FL to show the impacts of the number of selected devices, model aggregation error, and model broadcast error on the prediction accuracy. The detailed experiment setup is presented in Section IV.  To simulate the model distortion caused by model aggregation and broadcast errors, we add Gaussian noise to the received models at the BS and the devices, respectively. It is seen from Fig. \ref{Fig_MSE} that the performance gap between Benchmark (ideal situation without aggregation error and broadcast error) and the case without aggregation error is negligible. This is because broadcast error is applied to the initial local model, but the device can adjust its neural network to achieve a good training performance. However, if aggregation error exists, the effect of broadcast error aggravates the performance deterioration.  Thus, it is necessary to consider both aggregation error and broadcast error in the design of FL. Moreover, it is found that the number of selected devices is also a critical factor on the prediction accuracy of FL \cite{FL1}, \cite{FL2}, \cite{ML1}. Note that the errors are mainly caused by channel fading and additive noise during model aggregation and broadcast over wireless channels. Motivated by these observations, we propose to jointly conduct device selection and transceiver design to enhance the FL performance by maximizing the number of devices participating in FL while guaranteeing aggregation error and broadcast error to be within a certain range.

\subsection{Transmission Model}
As mentioned above, model broadcast and model aggregation are both implemented by parameter transmission over wireless channels. Referring to prior works on resource allocation for wireless FL, we apply the similar assumptions to facilitate analysis. First, we assume Rayleigh block flat-fading channels, where the channel coefficients remain constant within a communication block, but independently fade over successive blocks. Due to limited data storage and computing capability of intelligent devices, the dimension of model parameters $d$ is set to ensure that the entire model parameters can be transmitted within a communication block.  Moreover, it is assumed that the transmitted model parameters $\mathbf{q} \in \mathbb{R}^{d\times1}$ and $\mathbf{q}_i \in \mathbb{R}^{d\times1}, i \in \mathcal{K} $  are independent and normalized to have zero mean and unit variance \cite{YKFL,IRSFL}, i.e., $\mathbb{E}\left[\mathbf{q}\mathbf{q}^{\mathrm{H}}\right]=\mathbf{I}_{d}$ and $\mathbb{E}\left[\mathbf{q}_{i} \mathbf{q}_{i}^{\mathrm{H}}\right]=\mathbf{I}_{d}$. To simplify the notation, we will drop the communication round index $t$ and discuss a typical entry of the transmitted vector signal, which is denoted by $q$ at the BS for downlink and ${q}_i$ at the $i$-th device for uplink in the following.

\subsubsection{Model broadcast via downlink}
In the stage of model broadcast, the BS broadcasts the transmit signal $\mathbf{x}_{BS}$ constructed based on the global model parameters over the downlink channels as follows
\begin{equation}\label{eq7}
  \mathbf{x}_{BS}=\mathbf{w}q,
\end{equation}
where $\mathbf{w}$ is an $N$-dimensional transmit beam. Then, the received signal at the $k$-th device can be expressed as
 \begin{equation}\label{eq8}
  {{y}_{k}}=\mathbf{h}_{k}^{H}\mathbf{x}_{BS}+{{n}_{k}},
\end{equation}
where $\mathbf{h}_{k}$ denotes the channel vector from the $k$-th device to the BS, and ${{n}_{k}}$ is additive white Gaussian noise (AWGN) with variance $\sigma_0^2$. Therefore, the mean-square error (MSE) measuring the model distortion at the $k$-th device is given by
 \begin{eqnarray}\label{eq9}
  \!\!\!\!\text{MSE}_k^{\text{Device}}&=&\mathbb{E}\left\{ \left( {{v}_{k}}{{y}_{k}}-q \right){{\left( {{v}_{k}}{{y}_{k}}-q \right)}^{H}} \right\} \\ \nonumber
  &=&{{\left| {{v}_{k}}\mathbf{h}_{k}^{H}\mathbf{w}-1 \right|}^{2}}+\sigma _{0}^{2}{{\left| {{v}_{k}} \right|}^{2}},
\end{eqnarray}
where ${{v}_{k}}$ denotes the receiver scalar at the $k$-th device.

\subsubsection{Model aggregation via uplink}
\begin{figure*}[th] \centering
\includegraphics [width=0.9\textwidth] {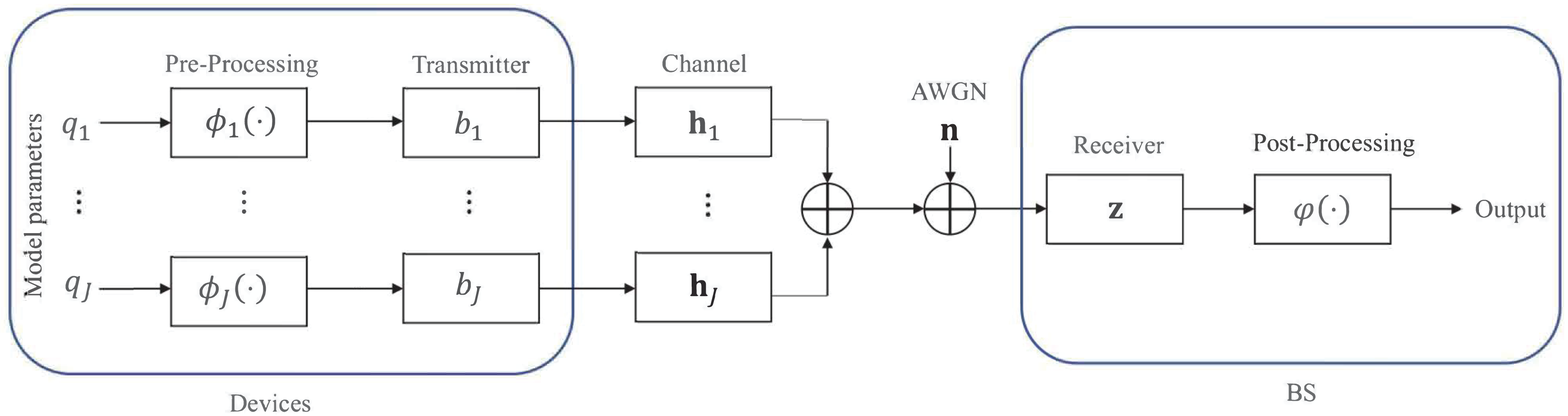}
\caption {The AirComp block diagram for the model aggregation.}
\label{Aircomp}
\end{figure*}

Since the average sum for model aggregation (\ref{eq1}) belongs to the category of nomographic functions \cite{Nomofun}, AirComp as a feasible solution is utilized to improve the communication efficiency for model aggregation from distributed devices. Based on the global loss function in (\ref{eq1}), let $\phi_i(x)=|D_i|x$ and $\varphi(x)=\frac{1}{\sum\limits_{i\in \mathcal{S}}{\left| {{D}_{i}} \right|}}x$ denote the pre-processing function at the $i$-th device and the post-processing function at the BS, c.f., Fig. \ref{Aircomp}, respectively. Thus, the target of aggregating local models at the BS is given by
\begin{equation}\label{eq10}
\tilde{q}=\frac{1}{\sum\limits_{i\in \mathcal{S}}{\left| {{D}_{i}} \right|}}\sum\limits_{i\in \mathcal{S}}{ {\left| {{D}_{i}} \right|}q_i},
\end{equation}
where $\mathcal{S}$ is the selected device set. To simplify the analysis and without loss of generality, we assume that all local datasets have a uniform size in the following, i.e., $|D_k|=D, \forall k\in \mathcal{K}$. In this context, the target of model aggregation in (\ref{eq10}) can be reduced as
\begin{equation}\label{eq10.5}
\tilde{q}=\frac{1}{\left| \mathcal{S} \right|}\sum\limits_{i\in \mathcal{S}}{{{q}_{i}}}.
\end{equation}
To minimize the distortion of the targeted function signal caused by channel fading and noise, it is desired to perform receive beamforming at the BS. Thus, the received signal at the BS is given by
\begin{equation}\label{eq11}
  \hat{q}={{\mathbf{z}}^{H}}\sum\limits_{i\in \mathcal{S}}{{{\mathbf{h}}_{i}}{{b}_{i}}{{q}_{i}}}+{{\mathbf{z}}^{H}}\mathbf{n},
\end{equation}
where $\mathbf{z}$ denotes the $N$-dimensional receive beam at the BS, $b_i$ is the transmit scalar at the $i$-th device, and $\mathbf{n}$ is the AWGN vector with the distribution $\mathcal{C} \mathcal{N}\left(\mathbf{0}, \sigma_{1}^{2} \mathbf{I}\right)$. Mathematically, the accuracy of model aggregation at the BS can be measured by the MSE between $\sum\limits_{i\in \mathcal{S}}{  {{q}_{i}} }$ and $\hat{q}$, which is given by
\begin{eqnarray} \label{eq12}
  \text{MSE}_{\text{BS}}&=&\mathbb{E}\left\{ \left( \hat{q}-\sum\limits_{i\in S}{{{q}_{i}}} \right){{\left( \hat{q}-\sum\limits_{i\in S}{{{q}_{i}}} \right)}^{H}} \right\} \\ \nonumber
  &=&\sum\limits_{i\in S}{{{\left| {{\mathbf{z}}^{H}}{{\mathbf{h}}_{i}}{{b}_{i}}-1 \right|}^{2}}}+\sigma _{1}^{2}{{\left\| \mathbf{z} \right\|}^{2}},
\end{eqnarray}

Checking (\ref{eq9}) and (\ref{eq12}), the broadcast error and aggregation error are jointly affected by the number of devices participating in FL, and the transceivers at the BS and the devices, and thus determine the FL performance. Hence, it makes sense to jointly perform device selection and transceiver design according to instantaneous CSI. In edge-intelligent networks, it is only able to obtain partial CSI due to imperfect channel estimation or feedback. Thus, it is necessary to take channel uncertainty into consideration for joint device selection and transceiver design, namely robust transmission scheme design. To characterize the channel uncertainty, we adopt a commonly-used deterministic CSI model \cite{CSI1,CSI2}. In particular, the real CSI $\mathbf{h}_i$ related to the $i$-th device can be modeled as
\begin{equation}\label{eq13}
  \mathcal{H}_{i} \triangleq\left\{\mathbf{h}_{i}=\hat{\mathbf{h}}_{i}+\mathbf{e}_{i} \mid\left\|\mathbf{e}_{i}\right\| \leq \varepsilon_{i}\right\},
\end{equation}
where $\hat{\mathbf{h}}_{i}$ is the obtained CSI and $\mathbf{e}_i$ is the CSI error, whose norm is within a given radius $\varepsilon_{i}$. In what follows, we make an effort to enhance the performance of FL from the perspective of efficient communication.

\section{Robust Design of Transmission Scheme for FL}
In this section, we aim to design a robust transmission scheme for communication-efficient FL in edge-intelligent networks.

\subsection{Problem Formulation}
Inspired by the key observations that the combination of the broadcast error over the downlink and the aggregation errors over the uplink may lead to a notable degradation of learning performance, while the convergence of training can be accelerated with more involved devices, we propose to simultaneously minimize the model distortion and maximize the number of involved devices \cite{YKFL}. In particular, we propose to maximize the number of selected devices while satisfying the MSE requirements of model broadcast via downlink and model aggregation via uplink, respectively. According to the characteristic of channel uncertainty in (\ref{eq13}), we attempt to design a worst-case robust transmission scheme, which is equivalent to the following optimization problem:
\begin{subequations}\label{OP1}
\begin{eqnarray}
\!\!\!\!\!\!\!\!\!\!\!\!\underset{\mathcal{S},{{v}_{i}},{{b}_{i}},\mathbf{w},\mathbf{z}}{\mathop{\max }}\,\!\!\!\!\!\!&&\!\!\!\!\left| \mathcal{S} \right| \label{OP1obj}\\
\!\!\!\!\textrm{s.t.}\!\!\!\!\!\!&&\!\!\!\!\!\!\underset{{{\mathbf{h}}_{i}}\in {\mathcal{H}_{i}}}{\mathop{\max }}\,\sum\limits_{i\in \mathcal{S}}{{{\left| {{\mathbf{z}}^{H}}{{\mathbf{h}}_{i}}{{b}_{i}}-1 \right|}^{2}}}+\sigma_{1}^{2}{{\left\| \mathbf{z} \right\|}^{2}}\le \gamma_B,\label{OP1st1}\\
\!\!\!\!&&\!\!\!\!\underset{{{\mathbf{h}}_{i}}\in {\mathcal{H}_{i}}}{\mathop{\max }}\,{{\left| {{v}_{i}}\mathbf{h}_{i}^{H}\mathbf{w}-1 \right|}^{2}}+\sigma _{0}^{2}{{\left| {{v}_{i}} \right|}^{2}}\le {{\gamma }_{D,i}}, i\in \mathcal{S}, \label{OP1st2}\\
\!\!\!\!&&\!\!\!\!{{b}_{i}}\le \sqrt{{{P}_{i}}},i\in \mathcal{S}, \label{OP1st3}\\
\!\!\!\!&&\!\!\!\!{{\left\| \mathbf{w} \right\|}^{2}}\le {{P}_{\max }},  \label{OP1st4}
\end{eqnarray}
\end{subequations}
where $\left| \mathcal{S} \right|$ in the objective function (\ref{OP1obj}) is the number of selected devices. $\gamma_B$ and ${{\gamma }_{D,i}}$ are the MSE requirements of model aggregation at the BS and model broadcast at the $i$-th device, respectively. (\ref{OP1st3}) and (\ref{OP1st4}) are the transmit power constraints at the $i$-th device with a power budget $P_i$ and at the BS with a power budget $P_{\max}$, respectively. It is intuitive that the mixed optimization problem (\ref{OP1}) is highly intractable due to the combinatorial objective function $\left| \mathcal{S} \right|$. To solve this challenge, we introduce an auxiliary variable $\bm{\chi}=[\chi_1,\chi_2,\ldots,\chi_K]^{T}$ to indicate the priority of being selected for devices, which characterizes the gap between the required MSE and the achievable MSE for the $k$-th device. In other words, the smaller the value of $\chi_k$, the higher priority the $k$-th device to be selected.  By adding $\chi_k$ to the right hand side of the corresponding MSE constraints, the objective function of maximizing the number of selected devices is equivalent to minimizing the number of non-zero $\chi_k$ (\cite{Convex}, section 11.4), where $\chi_k=0$ represents the $k$-th device being selected to participate in FL. Thus, problem (\ref{OP1}) can be reformulated as the following sparsity minimization problem:
\begin{subequations}\label{OP2}
\begin{eqnarray}
\!\!\!\!\!\!\!\!\!\!\!\!\!\underset{{{v}_{k}},{{b}_{k}},\mathbf{w},\mathbf{z},\bm{\chi},\rho_k,\vartheta_k}{\mathop{\min }}\,\!\!\!\!\!\!\!\!&&\!\!\!\!\|\bm{\chi}\|_0\label{OP2obj}\\
\!\!\!\!\textrm{s.t.}&&\!\!\!\!\underset{\|\mathbf{e}\|\le \varepsilon_k}{\mathop{\max }}\,{{{\left| {{\mathbf{z}}^{H}}{\left( {{{\mathbf{\hat{h}}}}_{k}}+{{\mathbf{e}}_{k}} \right)}{{b}_{k}}-1 \right|}^{2}}}\!\!\le\!\! \rho_k+\chi_k,  \label{OP2st1}\\
\!\!\!\!&&\!\!\!\!\sum\limits_{k\in \mathcal{K}}{\rho_k}+\sigma_{1}^{2}{{\left\| \mathbf{z} \right\|}^{2}}\!\!\le\!\! \gamma_B, \label{OP2st2} \\
\!\!\!\!&&\!\!\!\!\underset{\|\mathbf{e}\|\le \varepsilon_k}{\mathop{\max }}\,{{\left| {{v}_{k}}{\left( {{{\mathbf{\hat{h}}}}_{k}}+{{\mathbf{e}}_{k}} \right)}^{H}\mathbf{w}-1 \right|}^{2}}\le \vartheta_k, \label{OP2st3}\\
\!\!\!\!&&\!\!\!\!\vartheta_k+\sigma _{0}^{2}{{\left| {{v}_{k}} \right|}^{2}}\le {{\gamma }_{D,k}}+\chi_k, \label{OP2st4}\\
\!\!\!\!&&\!\!\!\!{{\left\| \mathbf{w} \right\|}^{2}}\le {{P}_{\max }} , \label{OP2st5}\\
\!\!\!\!&&\!\!\!\!{{b}_{k}}\le \sqrt{{{P}_{k}}}, \label{OP2st6}\\
\!\!\!\!&&\!\!\!\! \rho_k\geq0, \label{OP2st7}\\
\!\!\!\!&&\!\!\!\! \vartheta_k\geq0,\label{OP2st8}\\
\!\!\!\!&&\!\!\!\! \chi_k\geq0,\label{OP2st9}
\end{eqnarray}
\end{subequations}
where $\rho_k$ and $\vartheta_k$ are auxiliary variables for splitting the original MSE constraints in order to further deal with the non-convexity caused by channel uncertainty. However, the sparsity minimization problem (\ref{OP2}) is still nonconvex due to the nonconvex sparse objective function (\ref{OP2obj}) and nonconvex constraints (\ref{OP2st1}) and (\ref{OP2st3}) caused by channel uncertainty and coupled variables, i.e., transmitter $\mathbf{w}, b_i$ and receiver $\mathbf{z}, v_i$. To address these issues, we shall utilize the alternating optimization (AO) method \cite{AO} to decouple the problem and resort to some approximation means to solve the nonconvexity.

\subsection{Problem Solution}
In this section, we propose a two-step algorithm framework to solve problem (\ref{OP2}) for joint device selection and transceiver design, including sparsity inducing for determining the order of priority for devices and feasible detection for finding the maximum number of selected devices while ensuring the MSE requirements.

\subsubsection{Sparsity Inducing}
For the untractable and non-convex sparse objective function in the form of $\ell_0$-norm , we utilize a commonly used $\ell_1$-relaxation in the form of $\ell_1$-norm \cite{YKFL}, which is based on minimizing the sum of the infeasibilities \cite{Convex}. Thus, problem (\ref{OP2}) can be rewritten as
\begin{eqnarray}
\underset{{{v}_{k}},{{b}_{k}},\mathbf{w},\mathbf{z},\bm{\chi},\rho_k,\vartheta_k}{\mathop{\min }}\,&&\!\!\!\!\|\bm{\chi}\|_1 \label{OP3}\\
\textrm{s.t.}&&\!\!\!\!(\ref{OP2st1})-(\ref{OP2st9}),  \nonumber
\end{eqnarray}
Further, to address the non-convex constraint (\ref{OP2st1}) and (\ref{OP2st3}) involving the channel uncertainty, we have to introduce the following lemmas.

\emph{Lemma 1}: (Schurs's complement, \cite{Convex}) Let $\mathbf{D}$ be a Hermitian matrix defined as $\mathbf{D}\text{=}\left[ \begin{matrix}
   \mathbf{A} & {{\mathbf{B}}^{H}}  \\
   \mathbf{B} & \mathbf{C}  \\
\end{matrix} \right]$, where $\mathbf{D}$ is positive semi-definite if and only if $\mathbf{A}-{{\mathbf{B}}^{H}}{{\mathbf{C}}^{\text{-1}}}\mathbf{B}\succeq \mathbf{0}$ with the invertible matrix $\mathbf{C}$ or $\mathbf{C}-{{\mathbf{B}}^{H}}{{\mathbf{A}}^{\text{-1}}}\mathbf{B}\succeq \mathbf{0}$ with the invertible matrix $\mathbf{A}$.

\emph{Lemma 2}: Let us define a matrix function $\mathbf{F}\left( \mathbf{x} \right)=\mathbf{A}-{{\mathbf{B}}^{{H}}}\mathbf{x}{{\mathbf{c}}^{{H}}}-\mathbf{c}{{\mathbf{x}}^{{H}}}\mathbf{B}$, where variable vector $\mathbf{x}\in\mathbb{C}^{m\times 1}$, constant matrix $\mathbf{B}\in\mathbb{C}^{m\times n}$, constant vector  $\mathbf{c}\in\mathbb{C}^{n\times n}$, and constant Hermitian matrix $\mathbf{A}\in\mathbb{C}^{n\times n}$. Then,
\begin{equation}\nonumber
  \mathbf{F}\left( \mathbf{x} \right)\ge \mathbf{0},\forall \mathbf{x}:\left\| \mathbf{x} \right\|\le \varpi,
\end{equation}
holds true if and only if there exits $\lambda\geq0$, such that
\begin{equation}\nonumber
  \left[ \begin{matrix}
   \mathbf{A}-\lambda \mathbf{c}{{\mathbf{c}}^{H}} & -\varpi {{\mathbf{B}}^{H}}  \\
   -\varpi \mathbf{B} & \lambda {{\mathbf{I}}_{n}}  \\
\end{matrix} \right]\succeq \mathbf{0}.
\end{equation}
\begin{IEEEproof}
Please refer to Appendix A.
\end{IEEEproof}
Now, we first deal with the constraint (\ref{OP2st1}), i.e.,
\begin{eqnarray}\label{eq14}
  &&{{{\left| {{\mathbf{z}}^{H}}{\left( {{{\mathbf{\hat{h}}}}_{k}}+{{\mathbf{e}}_{k}} \right)}{{b}_{k}}-1 \right|}^{2}}} \nonumber \\
   &=&{{\left| \left( {{\mathbf{z}}^{H}}{{{\mathbf{\hat{h}}}}_{k}}{{b}_{k}}-1 \right)+{{\mathbf{z}}^{H}}{{\mathbf{e}}_{k}}{{b}_{k}} \right|}^{2}} \nonumber\\
  &\le& \rho_k+\chi_k.
\end{eqnarray}
Based on Lemma 1, (\ref{eq14}) can be transformed as
\begin{eqnarray}\label{eq15}
\!\!\!\!&&  \left[ \begin{matrix}
   {{\rho }_{k}}+{{\chi}_{k}} & b_{k}^{H}\mathbf{\hat{h}}_{k}^{H}\mathbf{z}-1+b_{k}^{H}\mathbf{ze}_{k}^{H}  \\
   {{b}_{k}}{{\mathbf{z}}^{H}}{{{\mathbf{\hat{h}}}}_{k}}-1+{{b}_{k}}{{\mathbf{z}}^{H}}{{\mathbf{e}}_{k}} & 1  \\
\end{matrix} \right]\succeq \mathbf{0}, \nonumber \\
\!\!\!\!&&\forall {{\mathbf{e}}_{k}}:\left\| {{\mathbf{e}}_{k}} \right\| \le {{\varepsilon }_{k}}.
\end{eqnarray}
Then, we let $  {{\mathbf{A}}_{k}}\triangleq \left[ \begin{matrix}
   {{\rho }_{k}}+{{\chi}_{k}} & {{b}_{k}^H}{{{\mathbf{\hat{h}}}}_{k}^{H}}{{\mathbf{z}}}-1  \\
   {{b}_{k}}{{\mathbf{z}}^{H}}{{{\mathbf{\hat{h}}}}_{k}}-1 & 1  \\
\end{matrix} \right],{{\mathbf{B}}_{k}}\triangleq \left[ \begin{matrix}
   \mathbf{0} & -b_{k}^{H}\mathbf{z}  \\
\end{matrix} \right],\mathbf{c}\triangleq {{\left[ \begin{matrix}
   1 & 0  \\
\end{matrix} \right]}^{T}}.$ Combine with (\ref{eq15}), we have
\begin{equation}\label{eq17}
  {{\mathbf{F}}_{k}}({{\mathbf{e}}_{k}})={{\mathbf{A}}_{k}}-\mathbf{B}_{k}^{H}{{\mathbf{e}}_{k}}{{\mathbf{c}}^{H}}-\mathbf{ce}_{k}^{H}{{\mathbf{B}}_{k}}\succeq \mathbf{0}, \forall {{\mathbf{e}}_{k}}:\left\| {{\mathbf{e}}_{k}} \right\|\le {{\varepsilon }_{k}}.
\end{equation}
According to Lemma 2, the constraint (\ref{OP2st1}) can be reformulated as
\begin{equation}\label{eq18}
\left[ \begin{matrix}
   {{\rho }_{k}}+{{\chi}_{k}}-{{\alpha }_{k}} & b_{k}^{H}\mathbf{\hat{h}}_{k}^{H}\mathbf{z}-1 & \mathbf{0}  \\
   {{b}_{k}}{{\mathbf{z}}^{H}}{{{\mathbf{\hat{h}}}}_{k}}-1 & 1 & {{\varepsilon }_{k}}{{b}_{k}}{{\mathbf{z}}^{H}}  \\
   \mathbf{0} & {{\varepsilon }_{k}}b_{k}^{H}\mathbf{z} & {{\alpha }_{k}}{{\mathbf{I}}_{N}}  \\
\end{matrix} \right]\succeq \mathbf{0},\exists {{\alpha }_{k}}\ge 0.
\end{equation}
Similarly, if there exists $\beta_k\ge0$, the constraint (\ref{OP2st3}) can be reformulated as
\begin{equation}\label{eq19}
  \left[ \begin{matrix}
   {{\vartheta }_{k}}-{{\beta }_{k}} & {{v}_{k}}\mathbf{\hat{h}}_{k}^{H}\mathbf{w}-1 & \mathbf{0}  \\
   {{\mathbf{w}}^{H}}{{{\mathbf{\hat{h}}}}_{k}}v_{k}^{H}-1 & 1 & \varepsilon _{k}\mathbf{w}{{v}_{k}}  \\
   \mathbf{0} & {{\varepsilon }_{k}} v_{k}^{H}{{\mathbf{w}}^{H}}& {{\beta }_{k}}\mathbf{I}_N  \\
\end{matrix} \right]\succeq \mathbf{0},\exists {{\beta }_{k}}\ge 0.
\end{equation}
Note that constraints (\ref{eq18}) and (\ref{eq19}) are still non-convex due to the coupled variables. To this end, we develop an AO method to decouple problem (\ref{OP3}). Specifically, problem (\ref{OP3}) can be divided into two subproblems, one for BS transceiver design and another for device transceiver design. The AO method will stop until the value of the objective function of the original problem approaches a stationary point in the iterations. In this context, the first subproblem of optimizing the device transceiver $\{v_k,b_k\}$ by given the BS transceiver $\{\mathbf{w,z}\}$ is equivalent to
\begin{subequations}\label{OP4}
\begin{eqnarray}
\!\!\!\!\!\!\underset{{{v}_{k}},{{b}_{k}},\bm{\chi},\rho_k,\vartheta_k,\alpha_k,\beta_k}{\mathop{\min }}\,&&\!\!\!\!\|\bm{\chi}\|_1 \\ \label{OP4obj}
\textrm{s.t.}&&\!\!\!\!(\ref{OP2st2}), (\ref{OP2st4}),  (\ref{OP2st6})-(\ref{OP2st9}),(\ref{eq18}),(\ref{eq19}),\nonumber \\
&&\!\!\!\! \alpha_k\geq0, \label{OP4st1} \\
&&\!\!\!\!\beta_k\geq0. \label{OP4st2}
\end{eqnarray}
\end{subequations}
It is seen that problem (\ref{OP4}) is a joint convex problem, and thus it can be solved by some off-the-shelf optimization toolboxes, e.g., CVX \cite{CVX}. Then, the second subproblem of optimizing the BS transceiver $\{\mathbf{w,z}\}$ with the obtained solution for device transceiver $\{v_k,b_k\}$  from problem (\ref{OP4}) can be expressed as
\begin{eqnarray}
\underset{\mathbf{w,z},\bm{\chi},\rho_k,\vartheta_k,\alpha_k,\beta_k}{\mathop{\min }}\,&&\!\!\!\!\|\bm{\chi}\|_1 \label{OP5} \\
\textrm{s.t.}&&\!\!\!\!(\ref{OP2st2}), (\ref{OP2st4}), (\ref{OP2st5}),(\ref{OP2st7})-(\ref{OP2st9}), \nonumber \\ &&\!\!\!\!(\ref{eq18}),(\ref{eq19}),(\ref{OP4st1}),(\ref{OP4st2}).\nonumber
\end{eqnarray}
It is observed that subproblem (\ref{OP5}) is also a joint convex problem, which can be solved by CVX directly. By iteratively solving the two subproblems, we can obtain the sparse vector $\bm{\chi}^{*}$. Next, we will conduct feasibility detection in the second step to find the feasible subset of devices $\mathcal{S}$  based on the obtained sparse vector.

\subsubsection{Feasibility Detection }
With the obtained sparse vector $\bm{\chi}$ from the above step, we first sort the elements $\chi_k, \forall k$ in an ascending order, i.e., ${{\chi}_{\pi \left( 1 \right)}}\le {{\chi}_{\pi \left( 2 \right)}}\le \cdots \le {{\chi}_{\pi \left( K \right)}}$, where ${{\chi}_{\pi \left( i \right)}}$ represents the $i$-th smallest element in $\{\chi_1,\cdots,\chi_K\}$. Then, we aim to find the maximum $m \in [1,K]$ that enables all devices in the selected devices set $\mathcal{S}^{[m]}=\{{\pi \left( 1 \right)},{\pi \left( 2\right)}, \cdots, {\pi \left( m \right)} \}$ to be feasible, i.e., satisfying the MSE requirements. Specifically, for a given device set $\mathcal{S}^{[m]}$, we check the feasibility via comparing the required MSE and the obtained MSE value based on the following problem:
\begin{subequations}\label{OP6}
\begin{eqnarray}
\!\!\!\!\!\!\!\!\!\!\!\!\!\!\!\!\underset{\mathbf{w},\mathbf{z},{{b}_{i}},{{v}_{i}}}{\mathop{\min }}\!\!\!\!\,\!\!\!\!&&\!\!\!\!\underset{\left\| {{\mathbf{e}}_{i}} \right\|\le {{\varepsilon }_{i}}}{\mathop{\max }}\,\!\!\!\!\sum\limits_{i\in {\mathcal{S}^{[m]}}}{{{\left| {{\mathbf{z}}^{H}}\left( {{{\mathbf{\hat{h}}}}_{i}}+{{\mathbf{e}}_{i}} \right){{b}_{i}}-1 \right|}^{2}}} +\sigma _{1}^{2}{{\left\| \mathbf{z} \right\|}^{2}} \label{OP6obj} \\
\!\!\!\!\!\!\textrm{s.t.}\!\!\!\!\!\!&&\!\!\!\!\underset{\left\| {{\mathbf{e}}_{i}} \right\|\le {{\varepsilon }_{i}}}{\mathop{\max }}\,{{\left| {{v}_{k}}{\left( {{{\mathbf{\hat{h}}}}_{k}}+{{\mathbf{e}}_{k}} \right)}^{H}\mathbf{w}-1 \right|}^{2}}+\sigma _{0}^{2}{{\left| {{v}_{i}} \right|}^{2}}\le {{\gamma }_{D,i}},\label{OP6st1} \\
\!\!\!\!&&\!\!\!\!{{\left\| \mathbf{w} \right\|}^{2}}\le {{P}_{\max }}, \label{OP6st2} \\
\!\!\!\!&&\!\!\!\! {{b}_{i}}\le \sqrt{{{P}_{i}}},\forall i\in {\mathcal{S}^{[m]}}.\label{OP6st3}
\end{eqnarray}
\end{subequations}
 Since the structure of nonconvex problem (\ref{OP6}) is similar to problem (\ref{OP3}), i.e., nonconvexity due to coupled variables and channel uncertainty, we just give the transformed convex subproblems directly. Specifically, the first subproblem of optimizing the device transceiver $\{v_i, b_i\}$ with the fixed BS transceiver $\{\mathbf{w,z}\}$ is given by
\begin{subequations}\label{OP7}
\begin{eqnarray}
\!\!\!\!\!\!\!\!\!\!\!\!\underset{v_i,b_i,\rho_i,\vartheta_i,\alpha_i,\beta_i}{\mathop{\min }}\!\!\!\!\!\!\!\!\,\!\!\!\!&&\!\!\!\! \sum\limits_{i\in {{S}^{[m]}}}{{{\rho }_{i}}}+\sigma _{1}^{2}{{\left\| \mathbf{z} \right\|}^{2}},\label{OP7obj} \\
\!\!\!\!\textrm{s.t.}\!\!\!\!&&\!\!\!\! (\ref{OP2st7}),(\ref{OP2st8}),(\ref{OP4st1}),(\ref{OP4st2}),(\ref{OP6st3}),  \nonumber\\
\!\!\!\!&&\!\!\!\!\left[ \begin{matrix}
   {{\rho }_{i}}-{{\alpha }_{i}} & b_{i}^{H}\mathbf{\hat{h}}_{i}^{H}\mathbf{z}-1 & \mathbf{0}  \\
   {{b}_{i}}{{\mathbf{z}}^{H}}{{{\mathbf{\hat{h}}}}_{i}}-1 & 1 & {{\varepsilon }_{i}}{{b}_{i}}{{\mathbf{z}}^{H}}  \\
   \mathbf{0} & {{\varepsilon }_{i}}b_{i}^{H}\mathbf{z} & {{\alpha }_{i}}{{\mathbf{I}}_{N}}  \\
\end{matrix} \right]\succeq \mathbf{0},\label{OP7st1} \\
\!\!\!\!&&\!\!\!\!  \left[ \begin{matrix}
   {{\vartheta }_{i}}-{{\beta }_{i}} \!\!&\!\! {{v}_{i}}\mathbf{\hat{h}}_{i}^{H}\mathbf{w}-1 \!\!&\!\! \mathbf{0}  \\
   {{\mathbf{w}}^{H}}{{{\mathbf{\hat{h}}}}_{i}}v_{i}^{H}-1 & 1 & \varepsilon _{i}\mathbf{w}{{v}_{i}}  \\
   \mathbf{0} & {{\varepsilon }_{i}} v_{i}^{H}{{\mathbf{w}}^{H}}& {{\beta }_{i}}\mathbf{I}_N  \\
\end{matrix} \right]\succeq \mathbf{0},\label{OP7st2} \\
\!\!\!\!&&\!\!\!\! \sum\limits_{i\in {{S}^{[m]}}}{{{\vartheta }_{i}}}+\sigma _{0}^{2}{{\left| {{v}_{i}} \right|}^{2}}\le {{\gamma }_{D,i}}. \label{OP7st3}
\end{eqnarray}
\end{subequations}
Then, the second subproblem of optimizing the BS transceiver $\{\mathbf{w,z}\}$ with the obtained device transceiver $\{v_i, b_i\}$ from (\ref{OP7}) can be expressed as
\begin{eqnarray}
\!\!\underset{\mathbf{w,z},\rho_i,\vartheta_i,\alpha_i,\beta_i}{\mathop{\min }}\!\!\!\!\,&&\!\!\!\! \sum\limits_{i\in {{S}^{[k]}}}{{{\rho }_{i}}}+\sigma _{1}^{2}{{\left\| \mathbf{z} \right\|}^{2}}, \label{OP8}\\
\textrm{s.t.}\!\!\!\!&&\!\!\!\! (\ref{OP2st7}),(\ref{OP2st8}),(\ref{OP4st1}),(\ref{OP4st2}),(\ref{OP6st2}),(\ref{OP7st1})-(\ref{OP7st3}). \nonumber
\end{eqnarray}
By alternatively solving these two subproblems, we can obtain a sub-optimal solution of problem (\ref{OP6}). If the obtained objective value is less than or equal to the required MSE $\gamma_B$ at the BS, then the set ${\mathcal{S}^{[m]}}$ can be chosen as a feasible set and corresponding aggregation error and broadcast error can be computed by designed transceivers. Otherwise, reset the selected devices set by letting $m=m-1$. In summary, a robust design of transmission scheme for FL in edge-intelligent networks can be described as Algorithm \ref{algrobust}.

\begin{algorithm}[h]
\caption{: Robust design of transmission scheme for FL in edge-intelligent networks.}
\label{algrobust}
\hspace*{0.02in} {\bf Input:}
$N,K,\sigma_0^2,\sigma_1^2,P_{\max}$,$\gamma, P_k,\varepsilon_k, \delta_k$ \\
\hspace*{0.02in} {\bf Output:}
$\mathcal{S}$, $ \text{MSE}_{\text{BS}}$, $\text{MSE}_{i}^{\text{Device}}$
\begin{algorithmic}[1]
\STATE{\textbf{Initialize} $\mathbf{z}^{[0]}, \mathbf{w}^{[0]}$, convergence accuracy $\nu_1$ and iteration index $t=1$;}
\REPEAT
 \STATE{obtain $\{v_k^{[t]},b^{[t]}_{k}\}$ by solving problem (\ref{OP4}) with fixed $\{\mathbf{w}^{[t-1]},\mathbf{z}^{[t-1]}\}$; }
 \STATE{obtain $\{ \mathbf{w}^{[t]},\mathbf{z}^{[t]},\bm{\chi}^{[t]} \}$ by solving problem (\ref{OP5}) with fixed $\{v_k^{[t]},b^{[t]}_{k}\}$;}
 \STATE{Compute $\triangle_1=|\|\bm{\chi}\|_1^{[t]}-\|\bm{\chi}\|_1^{[t-1]}|$}
  \STATE{$t=t+1$;}
\UNTIL{convergence: $\triangle_1<\nu_1$}
\STATE{obtain  $\bm{\chi}^{*}=\bm{\chi}^{[t]}$, and sort the elements by an ascending order, i.e., ${{\chi}_{\pi \left( 1 \right)}}\le {{\chi}_{\pi \left( 2 \right)}}\le \cdots \le {{\chi}_{\pi \left( K \right)}}$};
\STATE{\textbf{Initialize} the maximum number of selected devices $m=K$, convergence accuracy $\nu_2$; }
\WHILE{$m \geq 0$}
\STATE{Set the selected devices set $\mathcal{S}^{[m]}=\{{\pi \left( 1 \right)},{\pi \left( 2\right)}, \cdots, {\pi \left( m \right)} \}$;}
\REPEAT
 \STATE{ \textbf{Initialize} $\mathbf{z}^{[0]}, \mathbf{w}^{[0]}$, iteration index $t=1$}
 \STATE{Obtain $\{v_k^{[t]},b^{[t]}_{k}\}$ by solving problem (\ref{OP7}) with fixed $\{\mathbf{w}^{[t-1]},\mathbf{z}^{[t-1]}\}$; }
 \STATE{Obtain $\{ \mathbf{w}^{[t]},\mathbf{z}^{[t]}\}$ by solving problem (\ref{OP8}) with fixed $\{v_k^{[t]},b^{[t]}_{k}\}$;}
 \STATE{Compute $\triangle_2=| \text{MSE}_{\text{BS}}^{[t]}- \text{MSE}_{\text{BS}}^{[t-1]}|$}
  \STATE{$t=t+1$;}
\UNTIL{convergence: $\triangle_2 < \nu_2$}
\IF{$\text{MSE}_{\text{BS}}^{[t]} \leq \gamma$}
\STATE{$\mathbf{z}^{*}=\mathbf{z}^{[t]},\mathbf{w}^{*}=\mathbf{w}^{[t]}, v_i^{*}=v_i^{[t]}, b_i^{*}=b_i^{[t]}, \forall i \in \mathcal{S}^{[m]}$}
\STATE{\textbf{break;}}
\ENDIF
\STATE{$m=m-1$;}
\ENDWHILE
\STATE{Obtain $\mathcal{S}^{[m]}$, and compute $ \text{MSE}_{\text{BS}}$ and $\text{MSE}_{i}^{\text{Device}}$  according to the $\mathbf{z}^{*},\mathbf{w}^{*}, v_i^{*},b_i^{*}, \forall i \in \mathcal{S}^{[m]}$ .}
 \end{algorithmic}
\end{algorithm}

\emph{Optimality  Analysis:} It is observed that Algorithm \ref{algrobust} is divided into two main stages, one for sparse inducing (from step 1 to step 8) and another for feasibility detection (from step 9 to step 25). To overcome the non-convexity, we adopt well-known $\ell_1$-relaxation to approximate $\ell_0$-norm at the first stage and utilize the AO method to decouple the sub-problems in the overall process. As a result, the obtained solution might be not optimal. But the approximations will not shrink the feasible region too much over the original problem and the AO method will approach a local optimal point in the iterations.

%Then, for the stage of sparse inducing, since the transformed sub-problems (\ref{OP4}) and (\ref{OP5}) in terms of each optimization variable is convex, the optimal solutions to each sub-problem can be obtained by CVX for each iteration, which indicates the solutions in the $t$-th iteration are the feasible solution in the $(t+1)$-th iteration. That guarantees the objective value obtained in the $(t+1)$-th iteration is less than that in the $t$-th iteration, namely the sum of infeasiblities monotonically decreases after each iteration. Besides, owing to the constraints of transmit power at the BS and at the devices, the value of objective function is lower bounded. According to the monotone bounded convergence theorem, the proposed robust algorithm (from step 1 to step 8) for solving sparse inducing problem is convergent.  Analogously for the second stage, the transformed subproblems of feasibility detection (\ref{OP7}) and (\ref{OP8}) are both joint convex problems for each optimization variables, thus the objective value, namely aggregation error can be guaranteed monotonically decreases after each iteration. Thus, the proposed robust algorithm (from step 9 to step 25) for solving feasibility detection problem is also convergent. Moreover, we show the required number of iterations for solving sparse inducing problem under different MSE requirements at the BS by simulations in Fig. \ref{iterSparse} to verify that the proposed algorithm has a fast convergence.
\emph{Convergence  Analysis:} According to \cite{AO} and \cite{AO1}, the convergence of the AO algorithm is guaranteed when the solution of each subproblem in the iterations is its global optimum. As for the sparsity inducing problem (\ref{OP2}), we define its objective function at the $t$-iteration as $F_1(v_k^{[t]},b_k^{[t]},\mathbf{x}^{[t]},\mathbf{w}^{[t]},\mathbf{z}^{[t]})$. In the step 3 of Algorithm 2, since the transceivers at the devices can be optimally obtained for a given transceiver at the BS by solving the convex subproblem (\ref{OP4}), we have
\begin{equation}\label{ceq1}
  F_1(v_k^{[t]},b_k^{[t]},\mathbf{x}^{[t]},\mathbf{w}^{[t]},\mathbf{z}^{[t]})\geq F_1(v_k^{[t+1]},b_k^{[t+1]},\mathbf{x}^{[t]},\mathbf{w}^{[t]},\mathbf{z}^{[t]}).
\end{equation}
Then, in step 4 of Algorithm 2, the transceiver at the BS and the priority of being selected for devices can be optimally obtained for given transceivers at the devices by solving the convex subproblem  (\ref{OP5}), we have
\begin{eqnarray}\label{ceq2}
  &&F_1(v_k^{[t+1]},b_k^{[t+1]},\mathbf{x}^{[t]},\mathbf{w}^{[t]},\mathbf{z}^{[t]})\geq \nonumber \\
  &&F_1(v_k^{[t+1]},b_k^{[t+1]},\mathbf{x}^{[t+1]},\mathbf{w}^{[t+1]},\mathbf{z}^{[t+1]})
\end{eqnarray}
Based on (\ref{ceq1}) and (\ref{ceq2}), we can obtain
\begin{eqnarray}\label{ceq3}
  &&F_1(v_k^{[t]},b_k^{[t]},\mathbf{x}^{[t]},\mathbf{w}^{[t]},\mathbf{z}^{[t]})\geq  \nonumber \\ &&F_1(v_k^{[t+1]},b_k^{[t+1]},\mathbf{x}^{[t+1]},\mathbf{w}^{[t+1]},\mathbf{z}^{[t+1]}).
\end{eqnarray}
It is shown that the value of the objective function $F_1$ after each iteration of Algorithm 2 (step 1-8) is non-increasing. At the same time, the objective function value of problem (\ref{OP3}) has a lower bound, so the convergence of Algorithm 2 is guaranteed.  As for the feasibility detection problem (\ref{OP6}), we define its objective function at the $t$-iteration as $F_2(v_i^{[t]},b_i^{[t]},\mathbf{w}^{[t]},\mathbf{z}^{[t]})$ for a given device set. Similarly, since the subproblems (\ref{OP7}) and (\ref{OP8}) are both convex, we can obtain $F_2(v_k^{[t]},b_k^{[t]},\mathbf{w}^{[t]},\mathbf{z}^{[t]})\geq F_2(v_k^{[t+1]},b_k^{[t+1]},\mathbf{w}^{[t+1]},\mathbf{z}^{[t+1]})$. Thus, the proposed AO algorithm for solving the feasibility detection problem is also convergent. Moreover, based on the convergence rate analysis of AO algorithm in \cite{AO} and \cite{AO2}, it is known that the convergence rates of the proposed algorithms both show a two-stage behavior. Specifically, at first, the error decreases q-linearly until sufficiently small. After that, the sub-linear convergence is initiated.  Besides, we show the required number of iterations under various MSE requirements by simulations in Fig. \ref{iterSparse}. It is seen that Algorithm 2 has a fast convergence speed under different conditions, which also verifies the convergence property of Algorithm 2.

\emph{Complexity Analysis:}  As mentioned above, the proposed robust algorithm is split into two main stages, which have the similar structure in an iterative manner. Thus, we only discuss the per-iteration complexity in detail. It is seen that the main computational complexity of sparse inducing comes from the step 4, i.e., obtaining ${\{\mathbf{w,z},\bm{\chi}\}}$ by optimizing problem (\ref{OP5}) through a standard interior-point method (IPM) \cite{complexity}. Specifically, it has $2K$ linear matrix inequality (LMI) constraints of size $N+2$,  $6K$ LMI constraints of size 1, and $2$ second-order cone (SOC) constraints of size $N$. Thus, for a given precision $\Theta>0$ of solution, the per-iteration complexities of solving problem (\ref{OP5}) by IPM is $\ln \frac{1}{\Theta }\sqrt{10K+2KN+4}\cdot n\cdot \left[ 6K+2K{{\left( N+2 \right)}^{3}}+2{{N}^{2}}+n\left( n+6K+2K{{(N+2)}^{2}} \right) \right]$. Similarly,  the main computational complexity of solving the feasibility detection problem comes from the step 9, which has $2K$ LMI constraints of size $N+2$,  $5K$ LMI constraints of size 1, and $1$ SOC constraints of size $N$.  Thus, the per-iteration complexities of solving problem (\ref{OP7}) by IPM is $\ln \frac{1}{\Theta }\sqrt{9K+2KN+2}\cdot n\cdot \left[ 5K+2K{{\left( N+2 \right)}^{3}}+{{N}^{2}}+n\left( n+5K+2K{{(N+2)}^{2}} \right) \right]$, where the decision variable $n$ is on the order of $\mathcal{O}(KN^2)$.

\section{Simulation Results}
In this section, we provide extensive simulation results to validate the effectiveness and robustness of the proposed algorithm for FL in edge-intelligent networks.

\subsection{Performance on Parameter Transmission}
\begin{table*}
\small
\centering
\arrayrulecolor{black}
\caption{Simulation Parameters }\label{Simulation}
\begin{tabular}{|c|c|}
\hline
Parameters & Values \\ \hline
Number of BS antennas& $N=48$ \cite{NOMAbook}  \\\hline
Number of devices& $K=20$  \\\hline
CSI uncertainty  & $\varepsilon_{k}=\varepsilon=0.1$ \cite{CSI1}\\\hline
Maximum transmit SNR at the BS & $\text{SNR}_{B}=10 $ dB  \\\hline
Maximum transmit SNR at the devices & $\text{SNR}_{D,k}=\text{SNR}_{D}=10 $ dB  \\\hline
MSE requirement at the BS &  $\gamma_B=5$ dB \\\hline
MSE requirements at the devices &  $\gamma_{D,k}=\gamma_D=5$ dB\\
%\hline
%Noise Powers & $\sigma_0^2=-30$ dBm, $\sigma_1^2=-20$ dBm\\
\hline
\end{tabular}
\end{table*}

Without loss of generality, it is assumed that all devices are randomly distributed within a range of $100$ m from the BS.  To be close to the reality, the pass loss is modeled as $\mathrm{PL}_{\mathrm{dB}}=128.1+37.6\log_{10}(l)$ \cite{pathlossmodel}, where $l$ (km) is the distance between the BS and the device. For ease of analysis, we assume all devices have the same transmit power budget $P_k=P_0$ and the same MSE requirement $\gamma_{D,k}=\gamma_D$. Moreover,
 we use $\text{SNR}_B=10\log_{10}(P_{\max}/\sigma_0^2)$ to denote the transmit signal-to-noise ratio (SNR) (in dB) at the BS,  $\text{SNR}_{D}=10\log_{10}(P_{0}/\sigma_1^2)$ to denote the transmit SNR (in dB) at the device, $\gamma_{B}=10\log_{10}[\text{MSE}_{BS}/(\sigma_1^2/{P_{0}})]$ to denote the MSE requirement (in dB) at the BS, and $\gamma_{D,k}=10\log_{10}[\text{MSE}_k^{Device}/(\sigma_0^2/{P_{\max}})]$ to denote the MSE requirement (in dB) at the $k$-th device. The results averaged over 50 channel realizations.  Unless specified, the simulation parameters are set as in Table \ref{Simulation} following relevant standards \cite{CSI1}, \cite{NOMAbook}. It is worth pointing out that the proposed algorithm is applicable to various practical edge-intelligent networks by setting proper system parameters.

\begin{figure}[h]
\centering
\includegraphics [width=0.5\textwidth] {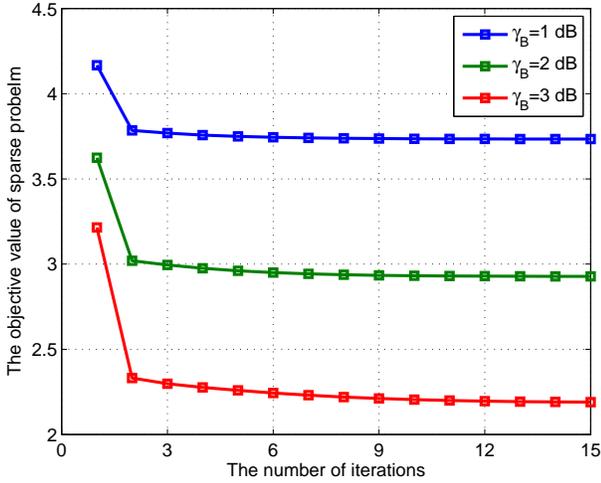}
\caption {Convergence behavior for sparse inducing of the proposed robust algorithm.}
\label{iterSparse}
\end{figure}
First, we present the convergence behavior of the proposed robust algorithm for solving sparse inducing problem using the AO method. As seen in Fig. \ref{iterSparse}, the objective value $\|\bm{\chi}\|_1$ decreases monotonically as expected, and converges to a stationary value within 10 iterations on average under different MSE requirements $\gamma_{B}$ at the BS, which verifies the proposed robust algorithm has a quick convergence speed. Moreover, it is seen that the smaller the value of $\gamma_B$, the faster the convergence, and the bigger the objective value. This is because that a small $\gamma_{B}$ represents a strict MSE requirement, leading to less sparsity and easy to find a minimum objective value.

\begin{figure}[h]
\centering
\includegraphics [width=0.5\textwidth] {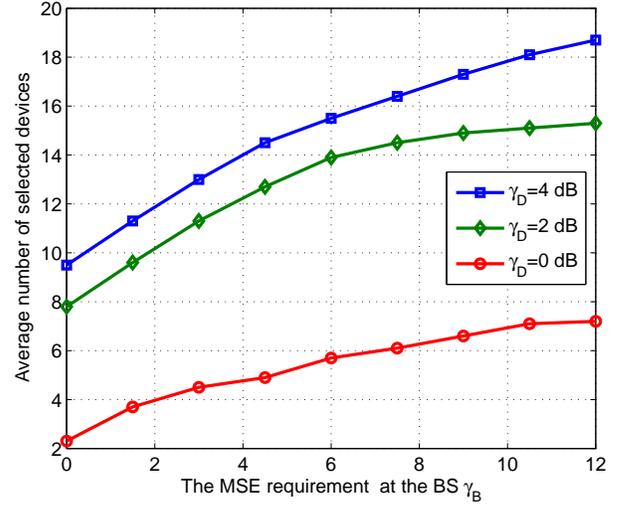}
\caption {Average number of selected devices versus the required MSE $\gamma_B$ at the BS for different MSE requirements $\gamma_D$ at the devices.}
\label{gammaB}
\end{figure}
Then, we show the impacts of the MSE requirements $\gamma_B$ for the BS and $\gamma_D$ for the devices on the performance of device selection for the proposed robust algorithm in  Fig. \ref{gammaB}. It is found that the average number of selected devices increases with the increment of the MSE requirements $\gamma_B$ for the BS. This is because that relaxing the condition of aggregation error at the BS can make more devices to be selected to participate in FL. Moreover, it is also found that the average number of selected devices can be increased by relaxing the MSE requirements at the devices. However, as mentioned above, the loose restrictions of MSE at the BS and at the devices mean the improvement of the distortion, leading to the performance degradation of the FL. Thus, it is desired to choose an appropriate number of devices under the suitable MSE requirements.

\begin{figure}[h]
\centering
\includegraphics [width=0.5\textwidth] {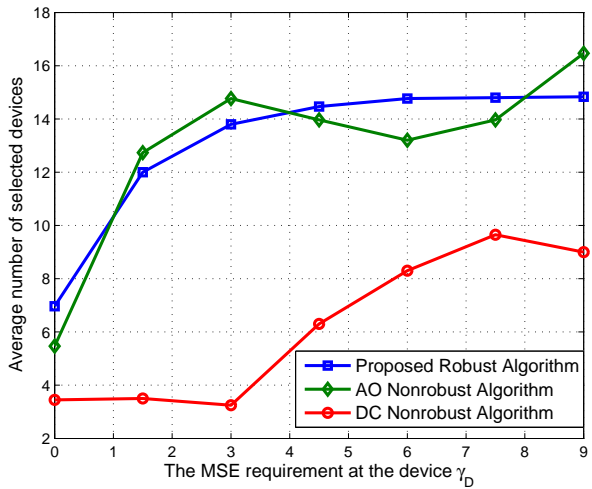}
\caption {Average number of selected devices versus the required MSE $\gamma_D$ at the devices for different device selection algorithms.}
\label{gammaD}
\end{figure}
Next, we compare the performance of different device selection algorithms. Note that the difference-of-convex (DC) nonrobust algorithm is designed based on the combination of \cite{parallel} and \cite{YKFL}. Specifically, for a given zero-forcing receiver $v_i$ at the device for downlink transmission \cite{parallel} and a zero-forcing transmitter $b_i$ at the device for uplink transmission \cite{YKFL}, the device selection and the transceiver at the BS are obtained based on the DC approach \cite{YKFL}. The AO nonrobust algorithm is the same as the proposed robust algorithm, i.e.,  adopting the AO method to design adaptive transceivers according to the system requirements, except that it does not take into account of channel uncertainty. It is seen that, for a given MSE requirement at the BS, the average number of selected devices increases as the required MSE at the device increases, but the improvement is finite. Moreover, for the two nonrobust algorithms, there seems to be no regularity in their trend. This is because they design the transceivers without considering the CSI uncertainty, sometimes selecting a few devices or even no device to ensure the MSE constraints. In particular, the proposed robust algorithm has the most stable performance, while the DC nonrobust algorithm has the worst performance. That quite confirms the effectiveness and robustness of the proposed robust algorithm.

\begin{figure}[h]
\centering
\includegraphics [width=0.5\textwidth] {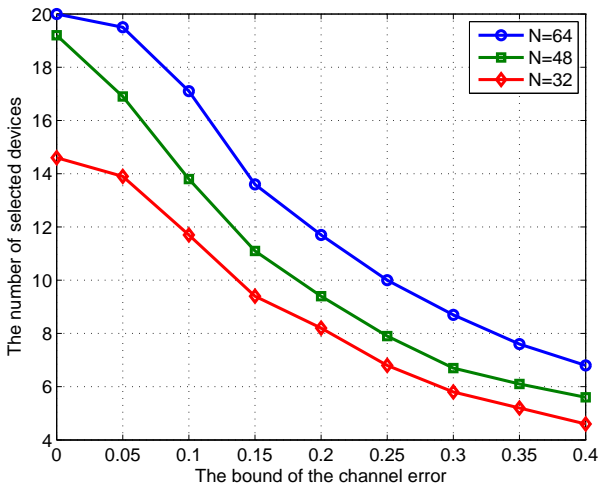}
\caption {Average number of selected devices versus the bound of the channel error $\varepsilon_0$ for different numbers of BS antennas.}
\label{gammaCompare}
\end{figure}
Fig. \ref{gammaCompare} investigates the effect of channel error bound $\varepsilon_0$ and the number of BS antennas $N$ on the performance of the proposed robust algorithm. Note that $\varepsilon_0=0$ represents the case of perfect CSI.  It is found that the average number of selected devices decreases as the bound of channel error increases. This is because a large channel error bound requires more resource to combat the channel uncertainty and guarantee the MSE in the worse case, and accordingly less resource contributes to improving the overall performance. Moreover, it is seen that the proposed robust algorithm with more BS antennas can select more devices to participate in FL due to more array gains for improving the performance. Especially, in the case of the number BS antennas $N=64$, the proposed robust algorithm selects all devices while ensuring MSE requirements. Thus, it is able to enhance the performance for the proposed robust algorithm by adding BS antennas.

\subsection{Performance on FL}
\begin{figure*}[!ht]
\centering
\includegraphics [width=0.8\textwidth] {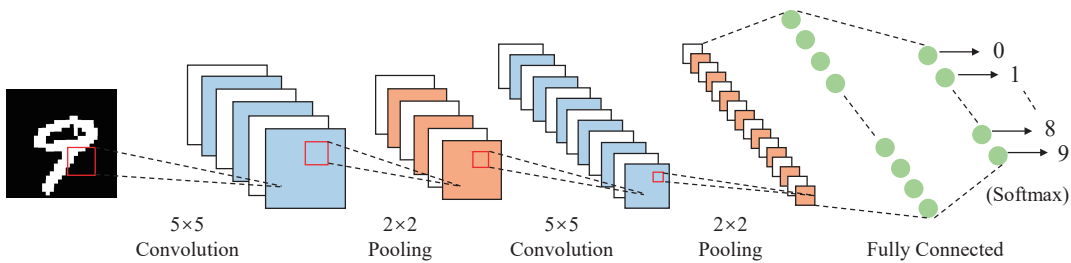}
\caption {Architecture of 6-layer CNN for training model used in experiments.}
\label{cnn}
\end{figure*}

\begin{figure*}[!ht]
\centering
\includegraphics [width=0.8\textwidth] {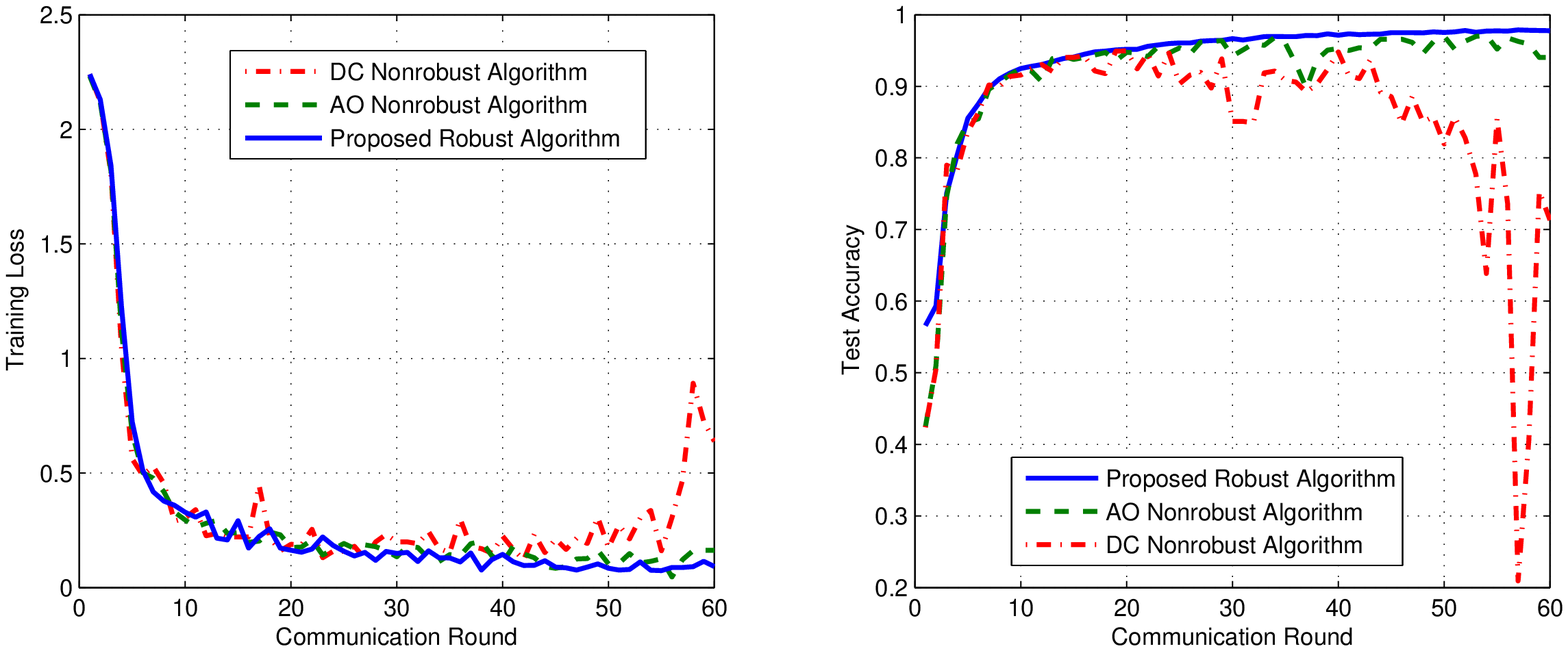}
\caption {FL Performance comparisons of different algorithms for parameter transmission with IID MNIST dataset.}
\label{iid}
\end{figure*}

\begin{figure*}[!ht]
\centering
\includegraphics [width=0.8\textwidth] {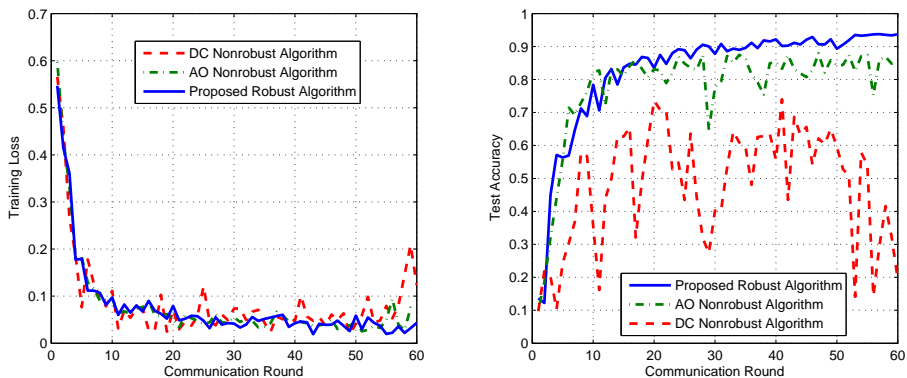}
\caption {FL Performance comparisons of different algorithms for parameter transmission with Non-IID MNIST dataset.}
\label{noniid}
\end{figure*}
In order to testify the performance of the proposed algorithm in practical FL applications, we consider a commonly used FL task of handwritten-digit recognition using the well-known MNIST dataset that consists of 10 categories ranging from digit ``0"  to ``9"  and a total of 70,000 labeled training data samples (60,000 for training and 10,000 for testing). To simulate the distributed mobile data, we consider two types of data partitions, i.e., the independent andidentically distributed (IID) setting and non-IID one. For the former setting, we randomly partition the training samples into 20 equal shards, each of which is assigned to one particular device. While for the latter setting, we first sort the data by digit label, divide it into 40 shards of 1,500 images, and randomly assign 2 shards to each device. As illustrated in Fig. \ref{cnn}, the training model is implemented using a 6-layer CNN that consists of two 5$\times$5 convolution layers with ReLu activation (the first with 32 channels, the second with 64 channels), each followed with 2$\times$2 max pooling layer, a fully connected layer with 512 units and ReLu activation, and a fully connected layer with Softmax classifier as the output layer (1,663,370 total parameters) \cite{FL3}. Each device updates its local model using the SGD algorithm with a learning rate being 0.01. The cross-entropy is used as the loss function, batchsize is 64, and epoch is 1. Note that different learning tasks, datasets, and training methods will have different prediction results. In particular, our adopted benchmark FL algorithm trained by such a 6-layer CNN on MNIST dataset for image classification presents over 95\% test accuracy in the ideal situation without aggregation error and broadcast error, cf. Fig. \ref{Fig_MSE}. Thus, it is fair and illustrates the performance of our proposed algorithm.

\begin{table*}[th]
  \caption{Some examples of handwritten-digit identification on IID MNIST dataset}
  \label{digit}
  \centering
  \begin{tabular}{ | c | c | c | c | c| c |c |c | }
    \hline
    Data image
    &\begin{minipage}[b]{0.07\columnwidth}
		\centering
		\raisebox{-.2\height}{\includegraphics[width=\linewidth]{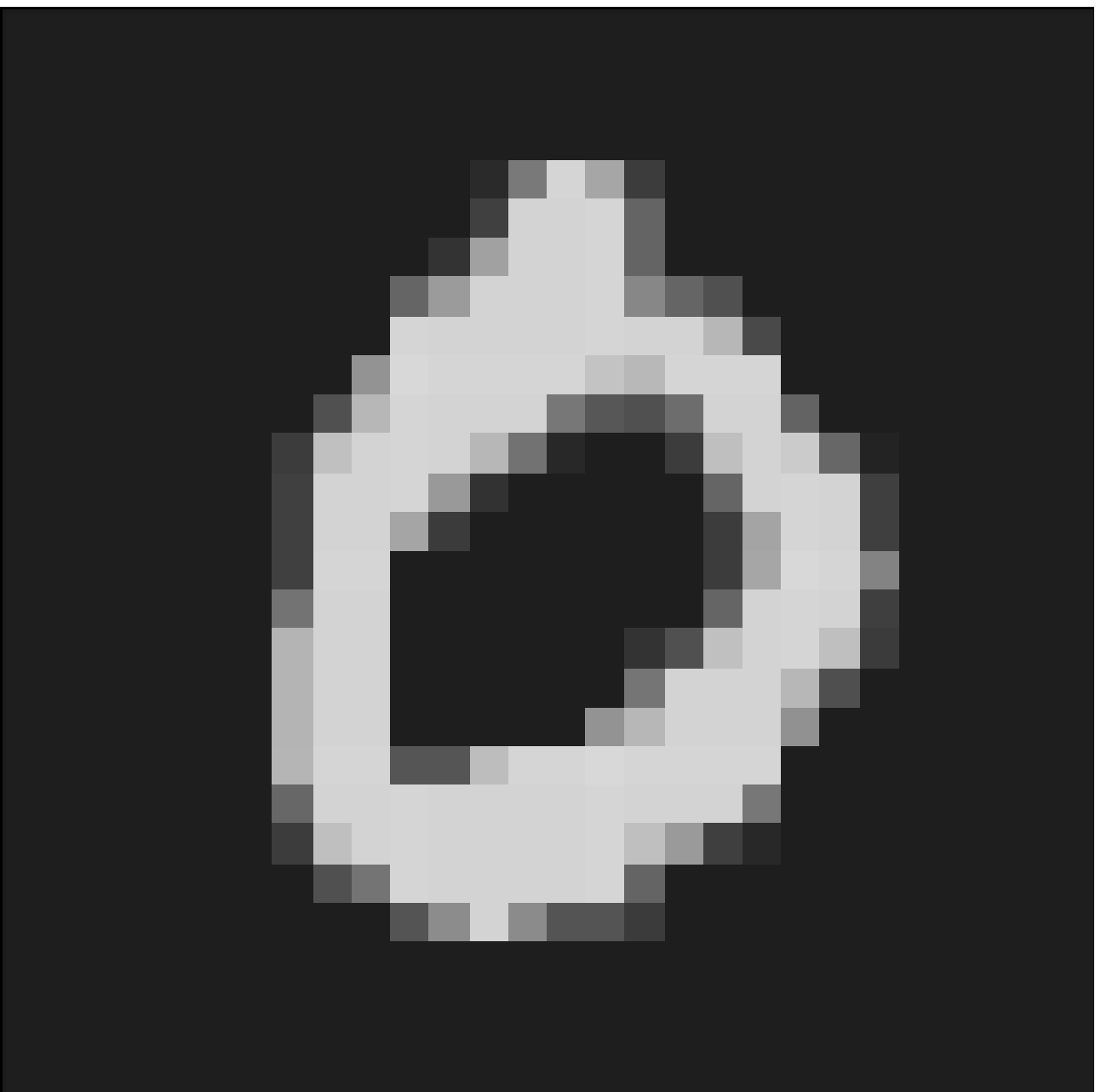}}
	\end{minipage}
    &     \begin{minipage}[b]{0.07\columnwidth}
		\centering
		\raisebox{-.2\height}{\includegraphics[width=\linewidth]{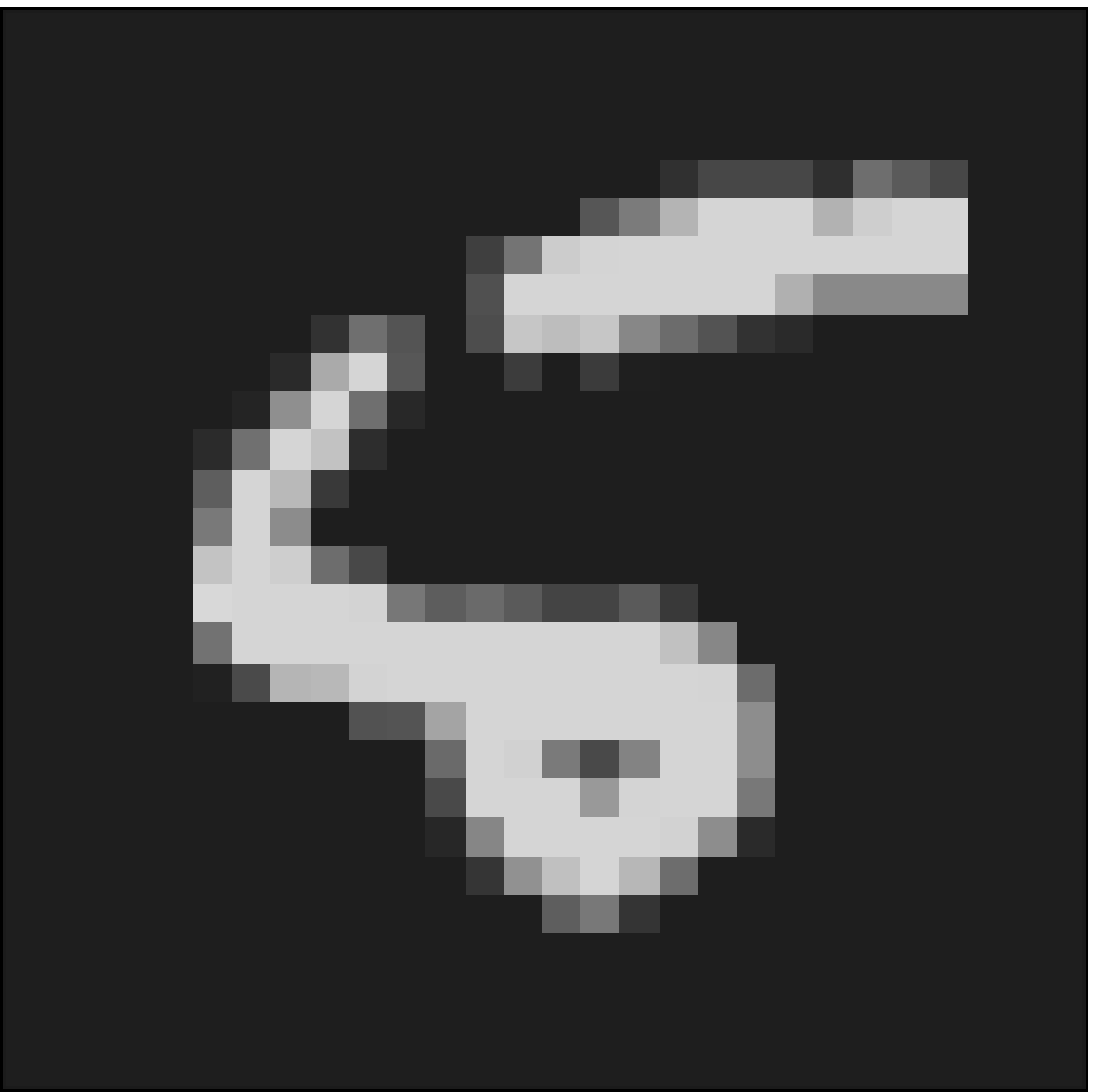}}
	\end{minipage}
    & \begin{minipage}[b]{0.07\columnwidth}
		\centering
		\raisebox{-.2\height}{\includegraphics[width=\linewidth]{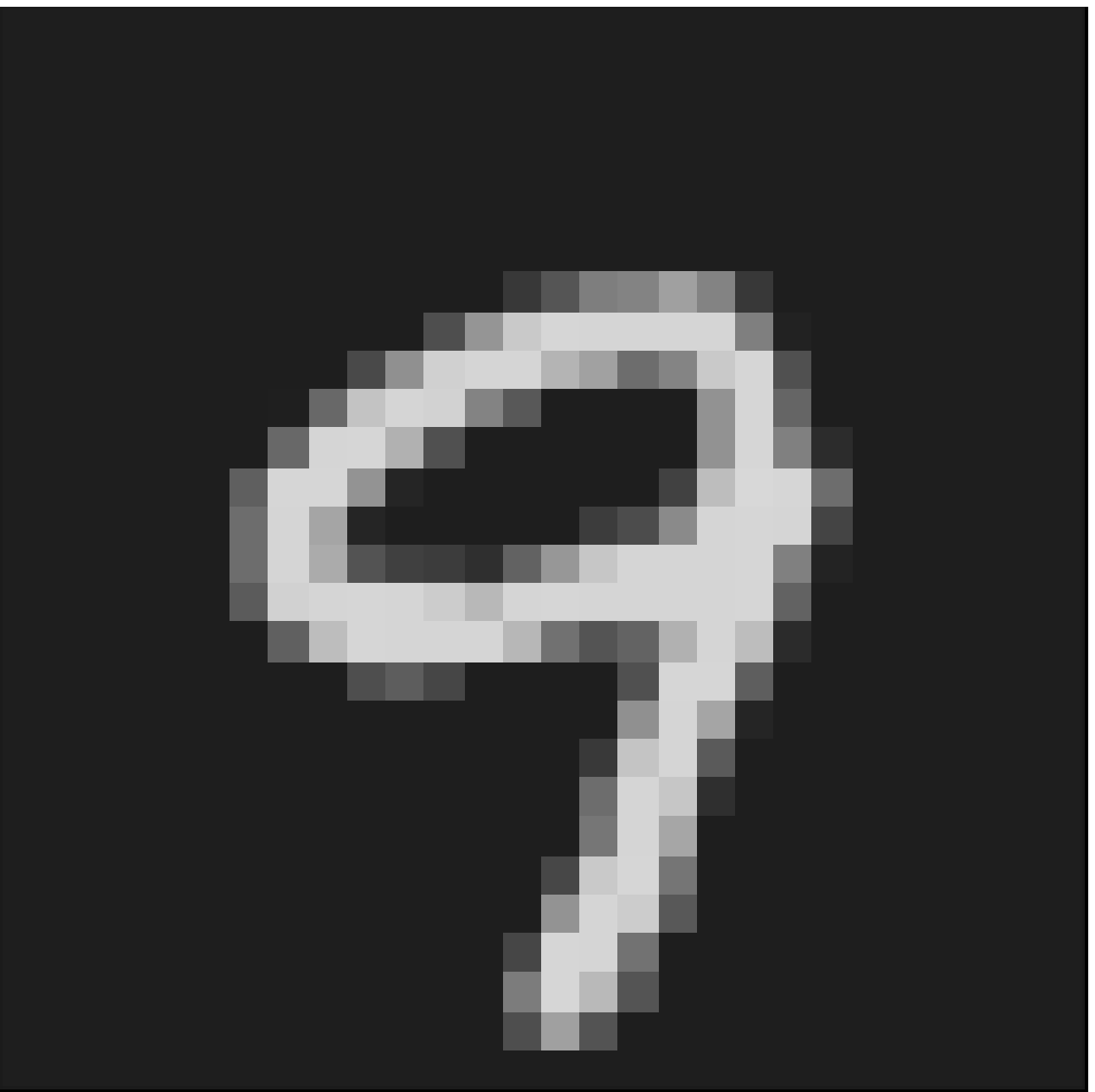}}
	\end{minipage}
& \begin{minipage}[b]{0.07\columnwidth}
		\centering
		\raisebox{-.2\height}{\includegraphics[width=\linewidth]{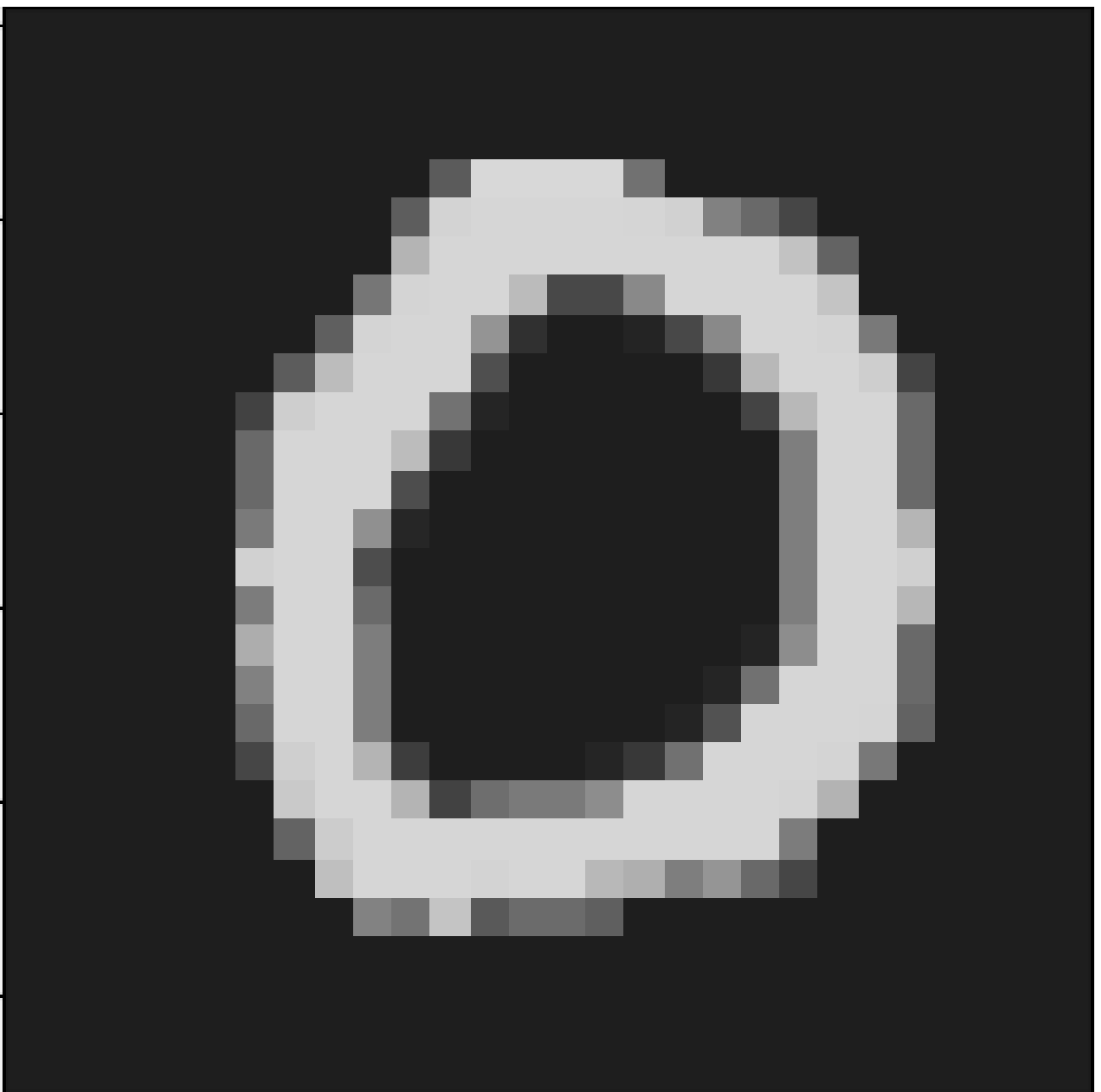}}
	\end{minipage}
& \begin{minipage}[b]{0.07\columnwidth}
		\centering
		\raisebox{-.2\height}{\includegraphics[width=\linewidth]{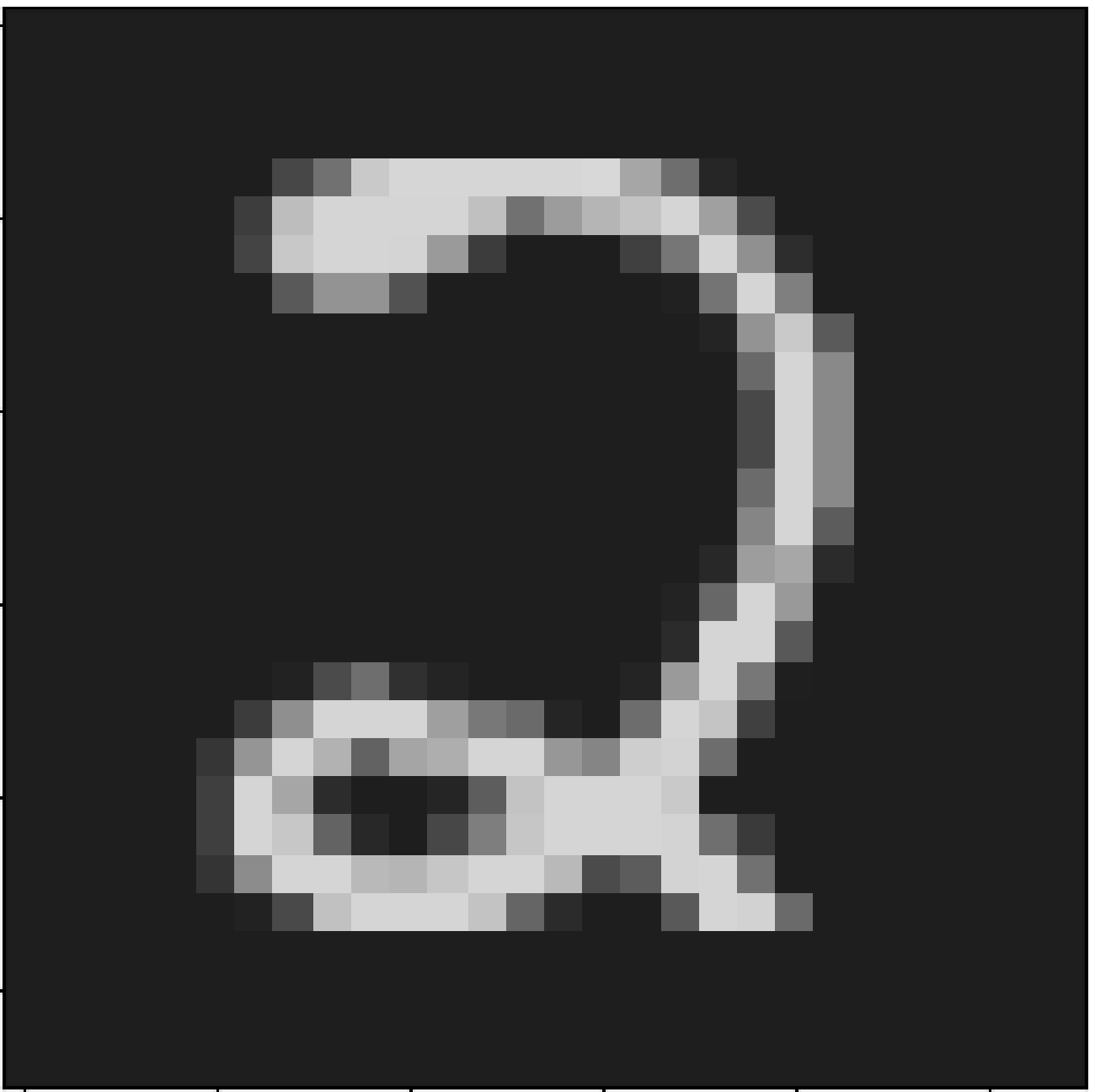}}
	\end{minipage}
& \begin{minipage}[b]{0.07\columnwidth}
		\centering
		\raisebox{-.2\height}{\includegraphics[width=\linewidth]{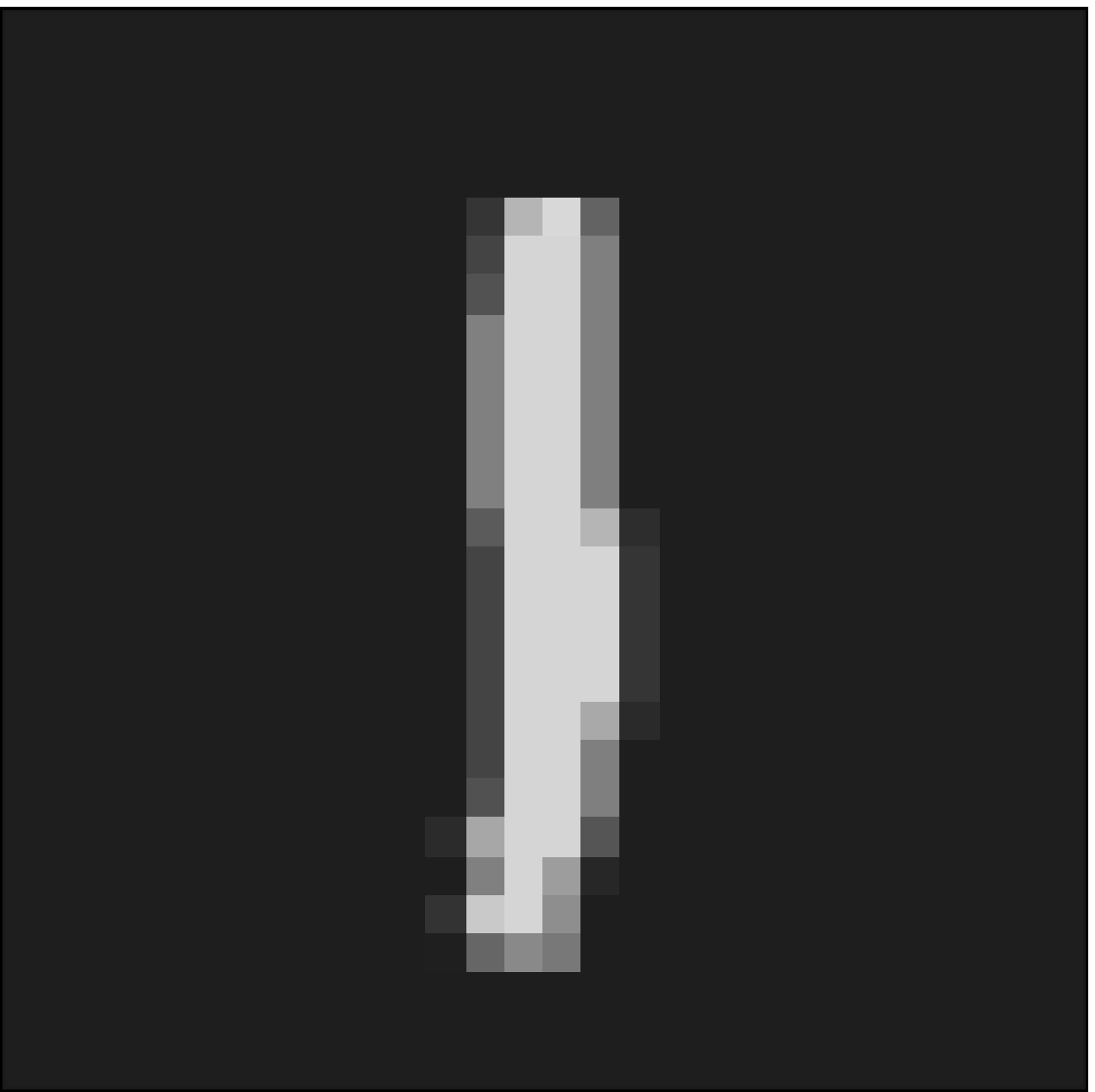}}
	\end{minipage}
& \begin{minipage}[b]{0.07\columnwidth}
		\centering
		\raisebox{-.2\height}{\includegraphics[width=\linewidth]{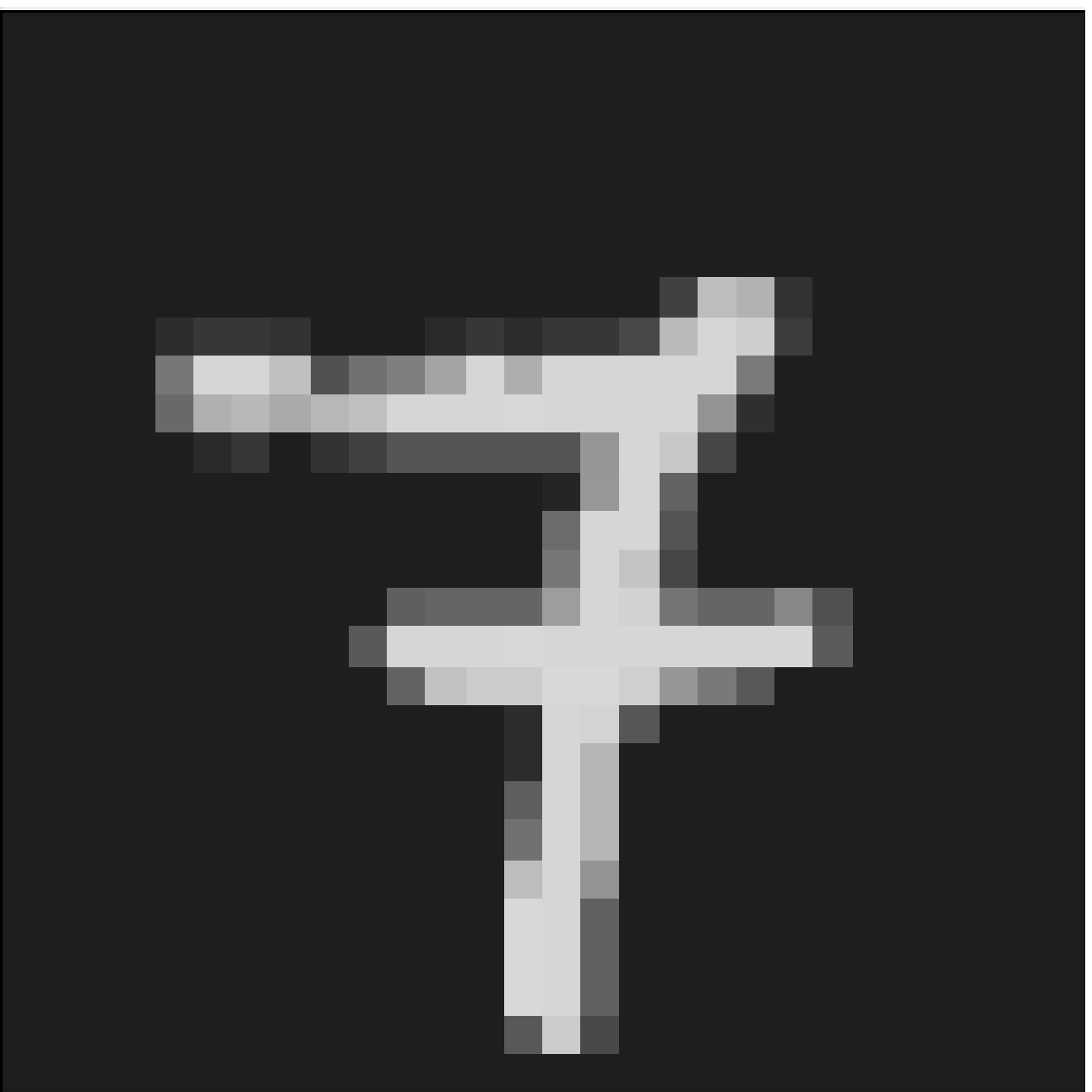}}
	\end{minipage}
\\ \hline
Proposed Algorithm &0 &5 & 9 &0 &2 & 1 &7   \\ \hline
AO Algorithm&0 &6($\times$)& 9& 0& 2& 1& 7 \\ \hline
DC Algorithm&0 &4($\times$) &4($\times$)& 0& 4($\times$)& 4($\times$) &4($\times$) \\ \hline\hline
Data image
& \begin{minipage}[b]{0.07\columnwidth}
		\centering
		\raisebox{-.2\height}{\includegraphics[width=\linewidth]{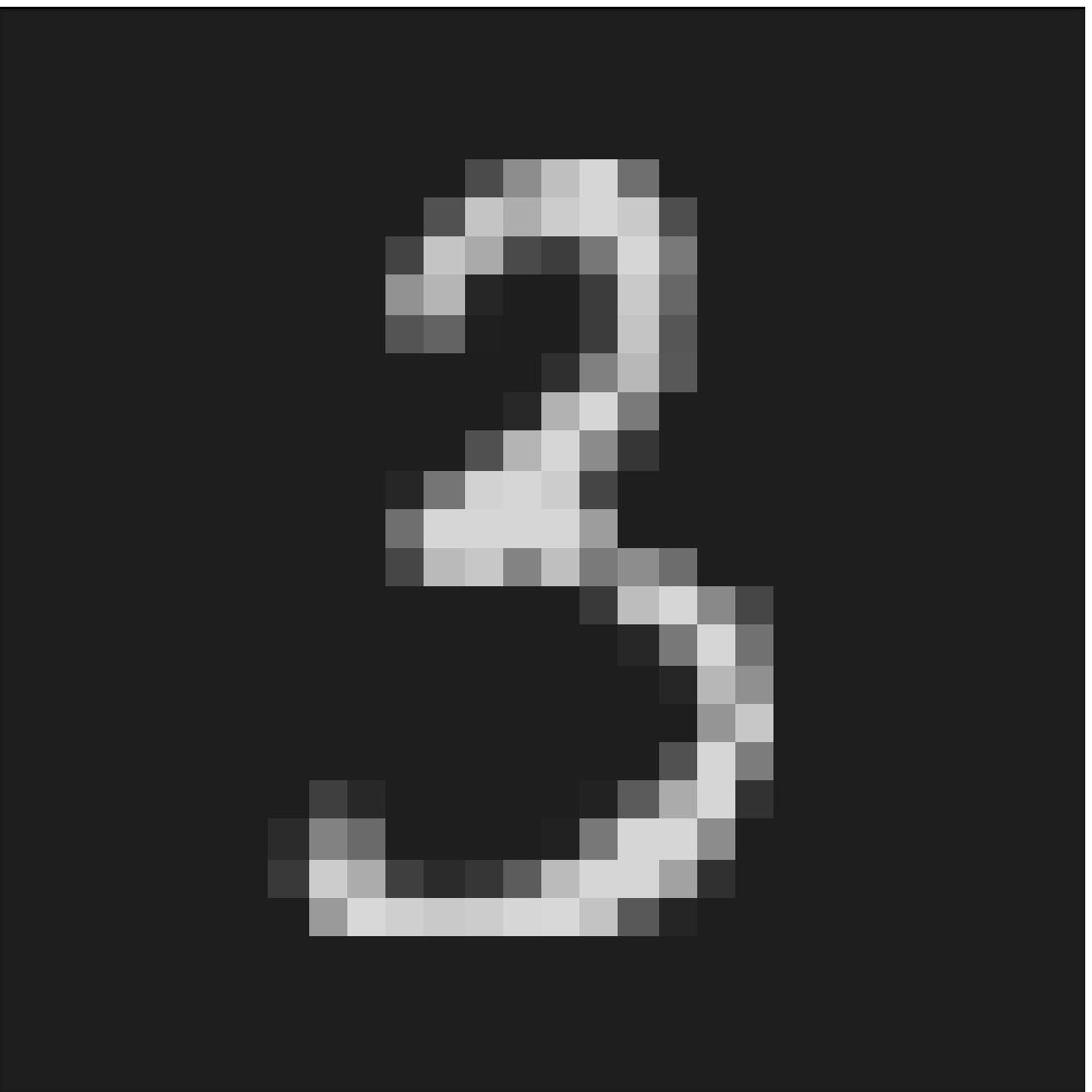}}
	\end{minipage}
& \begin{minipage}[b]{0.07\columnwidth}
		\centering
		\raisebox{-.2\height}{\includegraphics[width=\linewidth]{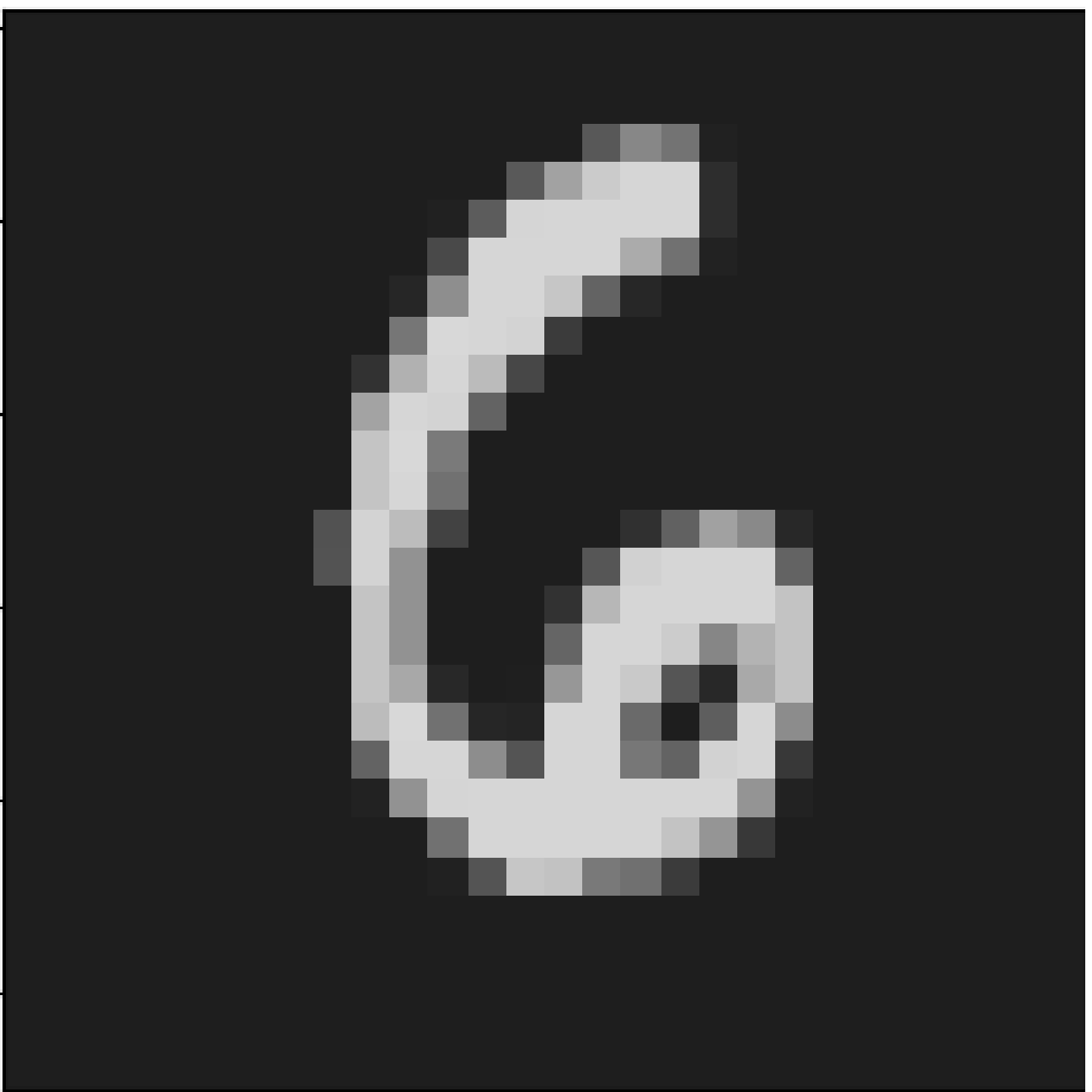}}
	\end{minipage}
& \begin{minipage}[b]{0.07\columnwidth}
		\centering
		\raisebox{-.2\height}{\includegraphics[width=\linewidth]{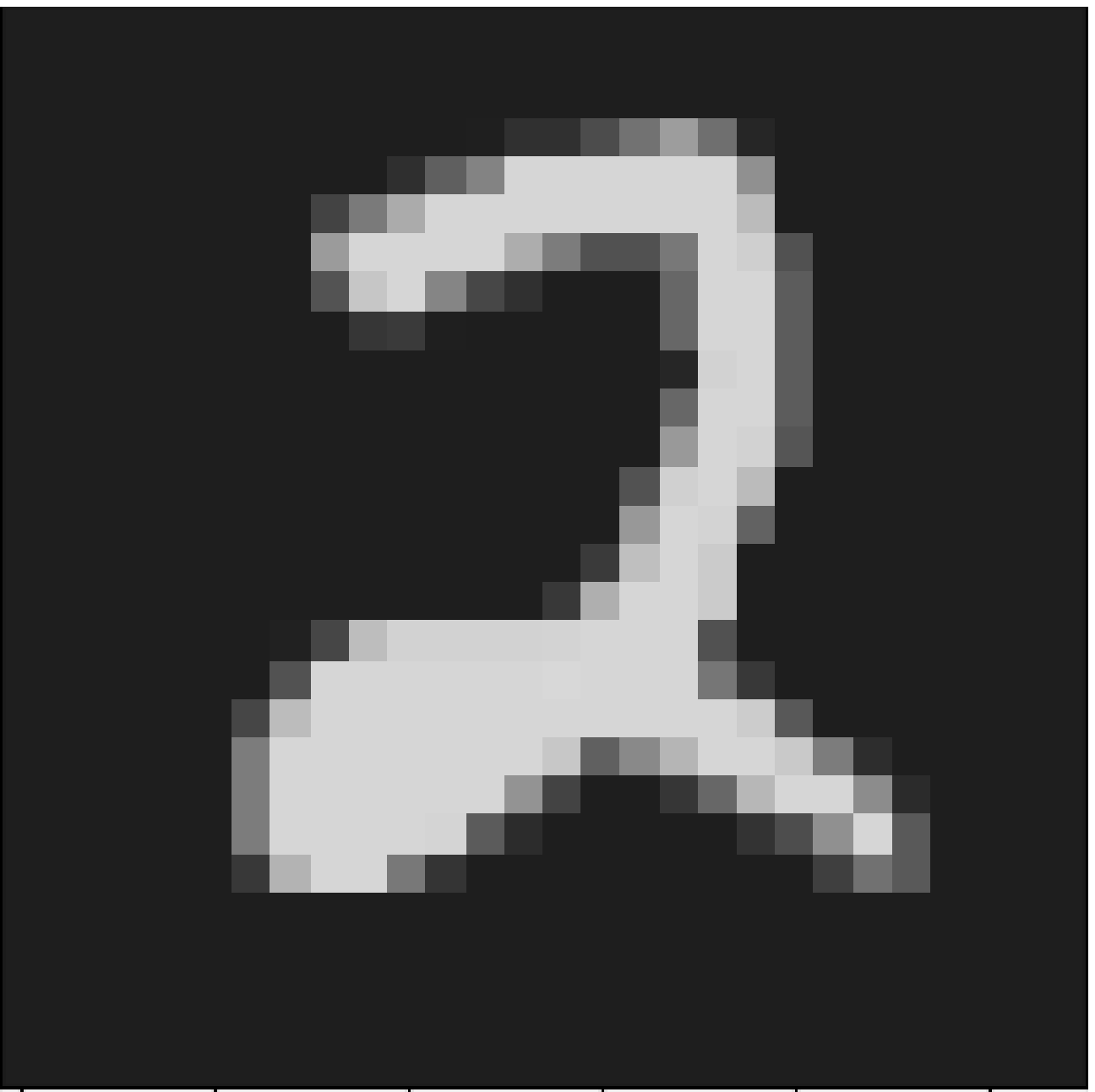}}
	\end{minipage}
& \begin{minipage}[b]{0.07\columnwidth}
		\centering
		\raisebox{-.2\height}{\includegraphics[width=\linewidth]{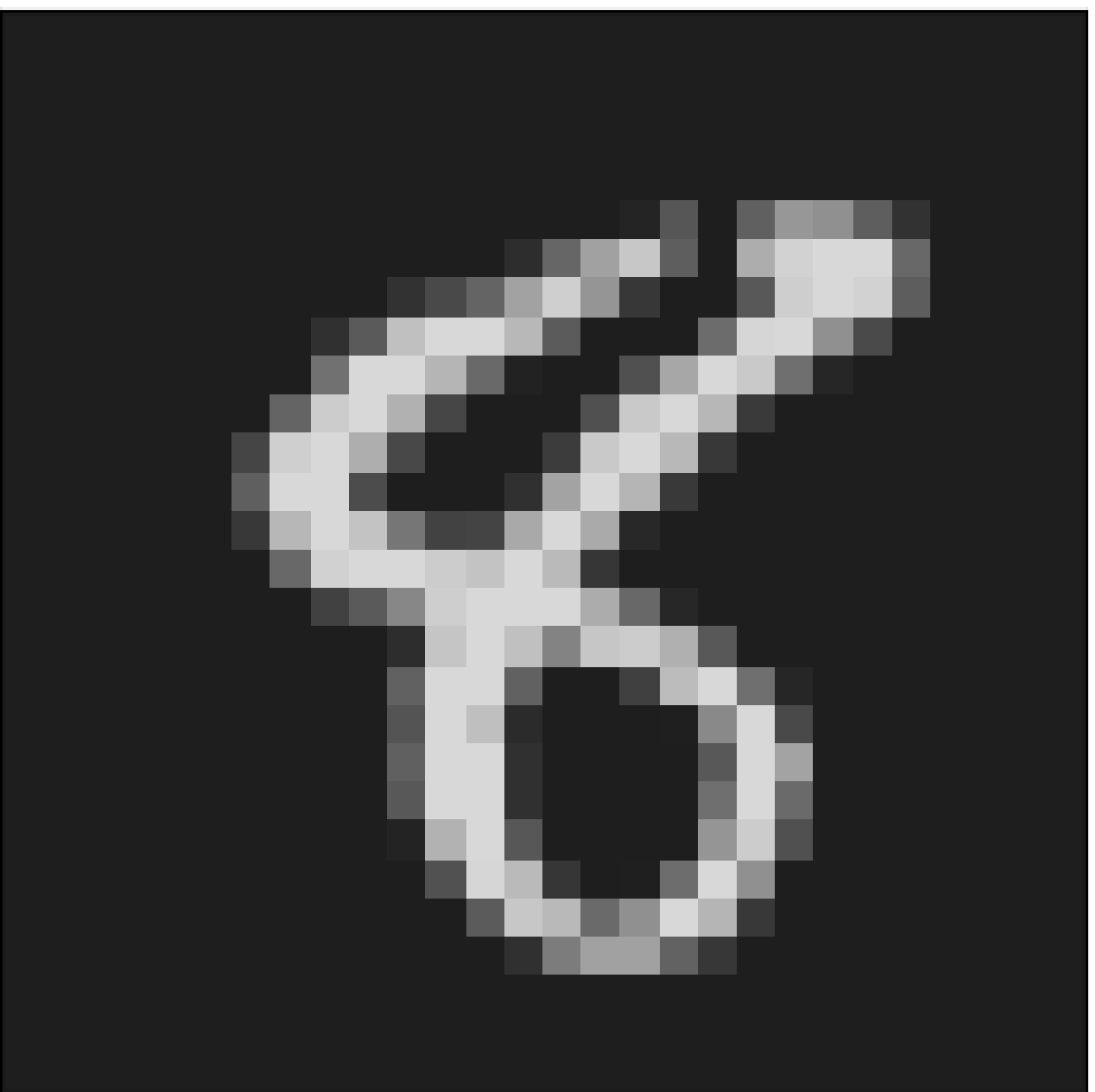}}
	\end{minipage}
& \begin{minipage}[b]{0.07\columnwidth}
		\centering
		\raisebox{-.2\height}{\includegraphics[width=\linewidth]{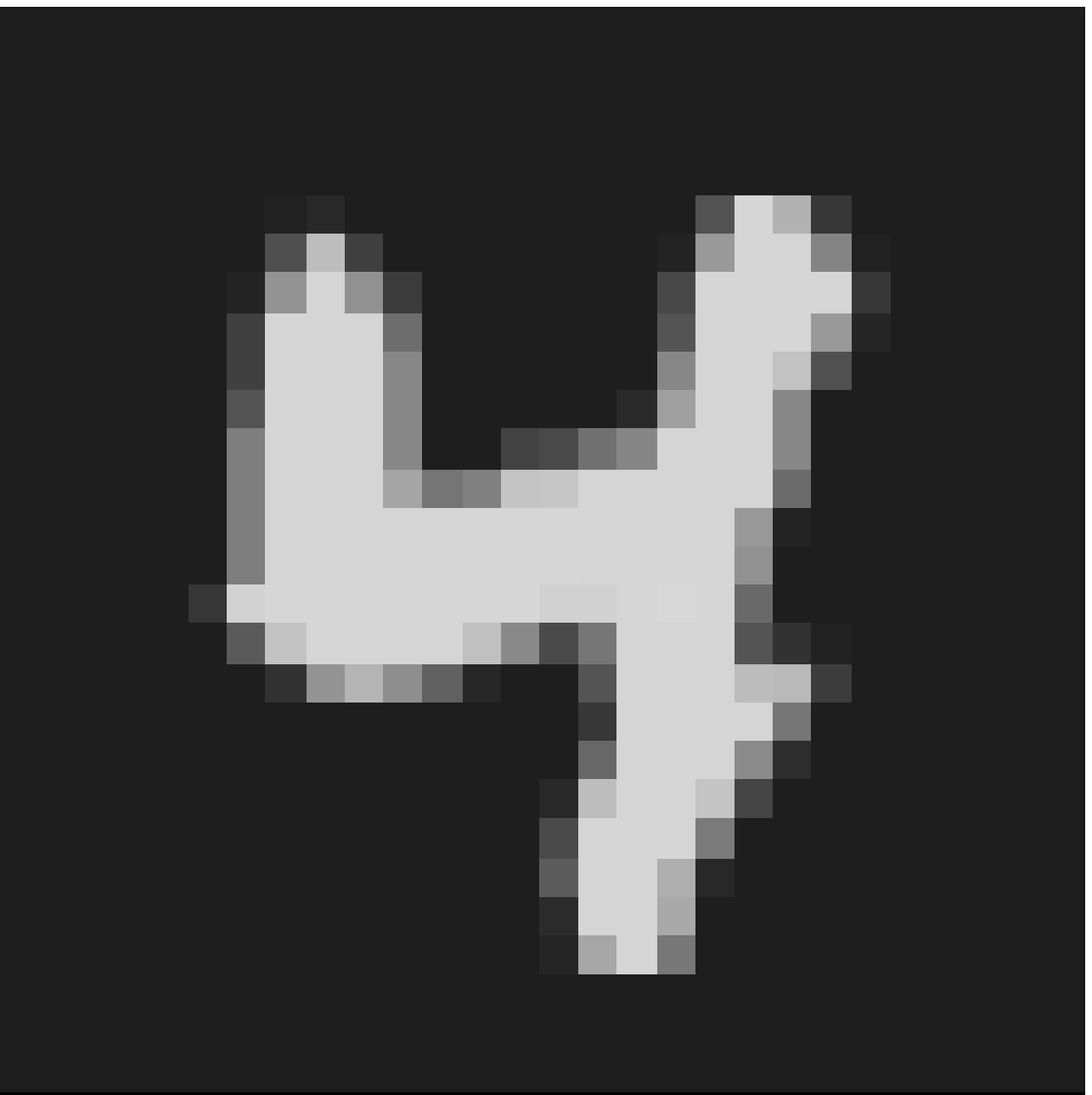}}
	\end{minipage}
& \begin{minipage}[b]{0.07\columnwidth}
		\centering
		\raisebox{-.2\height}{\includegraphics[width=\linewidth]{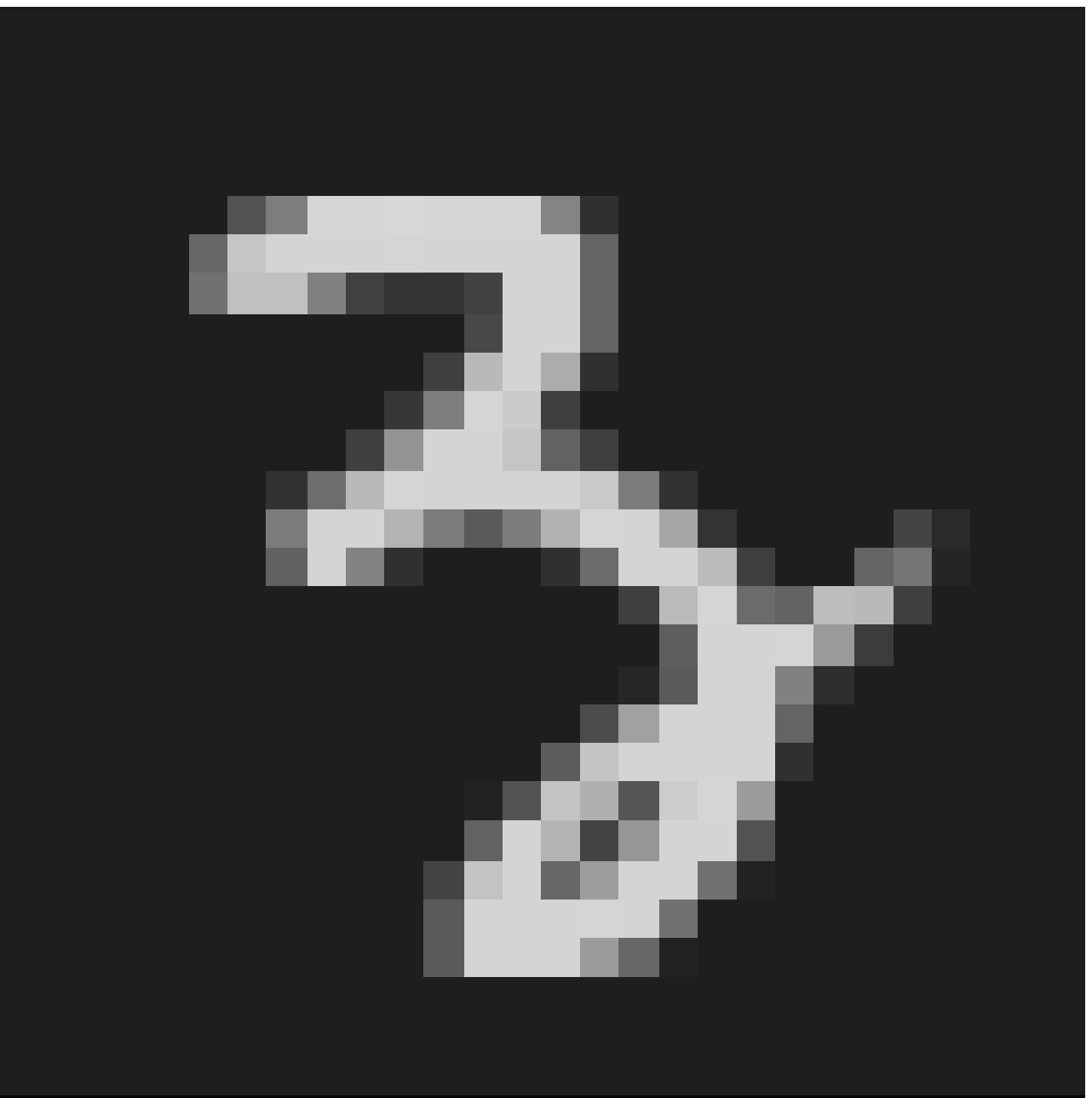}}
	\end{minipage}
& \begin{minipage}[b]{0.07\columnwidth}
		\centering
		\raisebox{-.2\height}{\includegraphics[width=\linewidth]{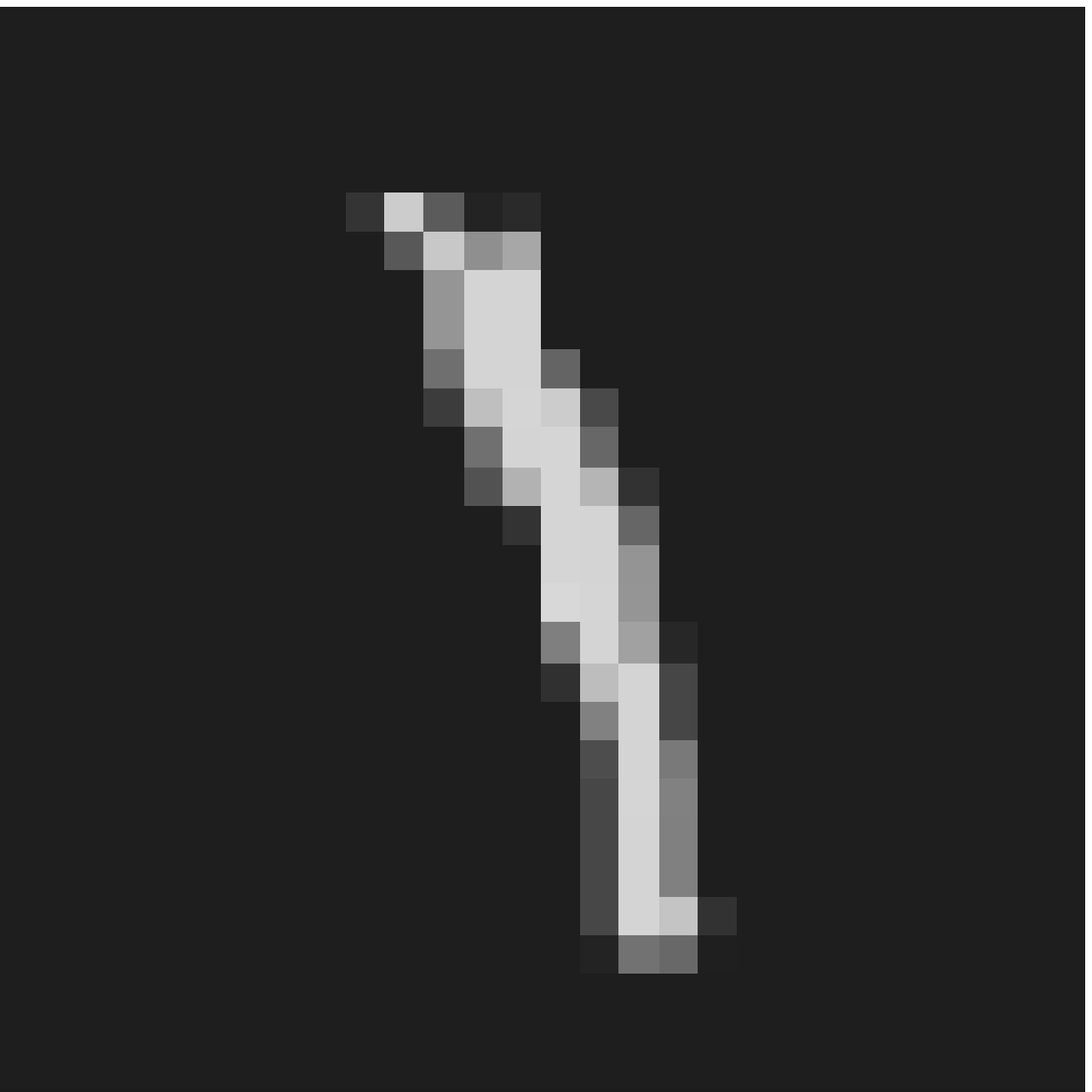}}
	\end{minipage}
    \\ \hline

Proposed Algorithm &3 &  6&  2&  8& 4& 3& 1    \\ \hline
AO Algorithm& 3 &6 &2& 8& 4& 3& 7($\times$) \\ \hline
DC Algorithm& 4($\times$)& 5($\times$) &4($\times$)& 4($\times$)& 4 &4($\times$)& 4($\times$)  \\ \hline
  \end{tabular}
\end{table*}

As shown in Fig. \ref{iid} and  \ref{noniid}, given $\text{SNR}_{B}=\text{SNR}_{D}=10 $ dB and  $\gamma_D=\gamma_B=5$ dB,  we compare the performance on the training loss and the test accuracy for the above three algorithms with the IID MNIST dataset and the Non-IID MNIST dataset, respectively. For the IID setting, it is seen that the proposed robust algorithm and the AO nonrobust algorithm both perform better than the DC nonrobust algorithm, and the proposed robust algorithm has a most stable and best performance on the training loss decreasing and the test accuracy increasing. For the Non-IID setting, it is found that the three algorithms all have the oscillatory waves on the performance. This is because some devices have not been selected to participate in FL due to the MSE constraints,  resulting in some data samples not being trained. For example, the $k$-th device is assigned two shards ``0" and ``9". If the $k$-th device is rarely selected in multiple communication rounds, then digit sample ``0" and ``9" are seldom trained, leading to a low prediction accuracy for digit sample "0" and "9". Even so, the proposed robust algorithm performs best and achieves a high test accuracy.  For vivid illustration, we present some examples of handwritten-digit identification in Table \ref{digit}.  These phenomenons confirm the effectiveness and robustness of the proposed robust algorithm for enhancing the performance of FL in edge-intelligent networks.

\section{Conclusion}
In this paper, we have investigated the factors affecting the accuracy of FL in edge-intelligent networks. For enhancing the performance of FL, we have proposed a robust parameter transmission scheme for both model broadcast and model aggregation, which has been formulated as an optimization problem by maximizing the number of selected devices participating in FL while ensuring the required MSE constraints on model broadcast and aggregation. To solve this problem, we have resorts to some approximation methods on convex transformation and proposed an AO-based robust algorithm including sparse inducing and feasibility detection. Numerical results have verified the effectiveness of the proposed algorithm for edge-intelligent networks. Moreover, it is found that whether or not the dataset is iid, the proposed robust algorithm can achieve a lower training loss and higher test accuracy compared with the nonrobust baselines.

\begin{appendices}
\section{The Proof of Lemma 2}
Prior to the proof, we provide a useful lemma in the following.

\emph{Lemma 3:} (S-procedure, \cite{Convex}) Let us consider a function ${{\mathbf{f}}_{m}}\left( \mathbf{x} \right)$ as
\begin{equation}\label{lemma3.1}
  {{\mathbf{f}}_{m}}\left( \mathbf{x} \right)={{\mathbf{x}}^{H}}{{\mathbf{A}}_{m}}\mathbf{x}+2\operatorname{Re}\left\{ \mathbf{b}_{m}^{H}\mathbf{x} \right\}+{{\mathbf{c}}_{m}},m\in \left\{ 1,2 \right\},\mathbf{x}\in {{\mathbb{C}}^{N\times 1}},
\end{equation}
where ${{\mathbf{A}}_{m}}\in {{\mathbb{C}}^{N\times N}},{{\mathbf{b}}_{m}}\in {{\mathbb{C}}^{N\times 1}}$ and ${{\mathbf{c}}_{m}}\in {{\mathbb{C}}^{N\times 1}}$. The derivation ${{\mathbf{f}}_{1}}\left( \mathbf{x}\right)\le 0\Rightarrow {{\mathbf{f}}_{2}}\left( \mathbf{x} \right)\le 0$ holds true if and only if there exists $\tau\geq0$, such that
\begin{eqnarray}\label{lemma3.2}
  \tau \left[ \begin{matrix}
   {{\mathbf{A}}_{1}} & {{\mathbf{b}}_{1}}  \\
   \mathbf{b}_{1}^{H} & {{\mathbf{c}}_{1}}  \\
\end{matrix} \right]-\left[ \begin{matrix}
   {{\mathbf{A}}_{2}} & {{\mathbf{b}}_{2}}  \\
   \mathbf{b}_{2}^{H} & {{\mathbf{c}}_{2}}  \\
\end{matrix} \right]\succeq \mathbf{0}.
\end{eqnarray}

\emph{Proof:} According to the definition of semi-positive matrix, if for $\forall \mathbf{d}\neq \mathbf{0}$, we have
\begin{equation}\label{proof1}
  {{\mathbf{d}}^{H}}\mathbf{Ad}-{{\mathbf{d}}^{H}}\left( {{\mathbf{B}}^{H}}\mathbf{x}\mathbf{c}^H+{{\mathbf{c}}}{{\mathbf{x}}^{H}}\mathbf{B} \right)\mathbf{d}\ge 0.
\end{equation}
 That confirms that $\mathbf{F}\left( \mathbf{x} \right)\succeq \mathbf{0},\text{   }\forall \mathbf{x}:\left\| \mathbf{x} \right\|\le \varpi \nonumber$ holds true. Based on Lemma 1, (\ref{proof1}) is equivalent to
\begin{eqnarray}\label{proof2}
  {{\mathbf{d}}^{H}}\mathbf{Ad}&\ge& \underset{{{\left\| \mathbf{x} \right\|}}\le \varpi }{\mathop{\max }}\,{{\mathbf{d}}^{H}}\left( {{\mathbf{B}}^{H}}\mathbf{xc}^H+{{\mathbf{c}}}{{\mathbf{x}}^{H}}\mathbf{B} \right)\mathbf{d}\nonumber\\
  &=&2\varpi | \mathbf{c}^H\mathbf{d}|\left\| \mathbf{Bd} \right\|.
\end{eqnarray}
By utilizing the \emph{Cauchy-Schwarz inequality} \cite{inequation}, (\ref{proof2}) can be further transformed as
\begin{equation}\label{proof3}
  {{\mathbf{d}}^{H}}\mathbf{Ad}-2\varpi \xi\mathbf{c}^H\mathbf{d}\ge 0,\forall \xi: |\xi| \le \left\| \mathbf{Bd} \right\|.
\end{equation}
Then, by exploiting Lemma 3 with $\varpi^2-{{\mathbf{d}}^{H}}{{\mathbf{B}}^{H}}\mathbf{Bd}\le 0$, (\ref{proof3}) is satisfied if and only if there exists a $\lambda \geq 0 $ such that
\begin{equation}\label{proof4}
  \left[ \begin{matrix}
   \mathbf{A}-\lambda {{\mathbf{c}}}\mathbf{c}^{H} & -\varpi {{\mathbf{B}}^{H}}  \\
   -\varpi \mathbf{B} & \lambda {{\mathbf{I}_n}}  \\
\end{matrix} \right]\succeq \mathbf{0}. \nonumber
\end{equation}
The proof is completed.

\end{appendices}


\begin{thebibliography}{1}


\bibitem{6G1}
P. Yang, Y. Xiao, M. Xiao, and S. Li, ``6G wireless communications: Vision and potential techniques," \emph{IEEE Netw.}, vol. 33, no. 4, pp. 70-75, Jul. 2019.

\bibitem{6G2}
W. Saad, M. Bennis, and M. Chen, ``A vision of 6G wireless systems: Applications, trends, technologies, and open research problems," \emph{IEEE Netw.}, vol. 34, no. 3, pp. 134-142, May 2020.


\bibitem{6G3}
Y. Chen, P. Zhu, G. He, X. Yan, H. Baligh, and J. Wu, ``From connected people, connected things, to connected intelligence," in \emph{Proc. 2020 6G SUMMIT}, Levi, Finland, 2020, pp. 1-7.


\bibitem{6G4}
H. Yang, A. Alphones, Z. Xiong, D. Niyato, J. Zhao, and K. Wu, ``Artificial-intelligence-enabled intelligent 6G networks," \emph{IEEE Netw.}, vol. 34, no. 6, pp. 272-280, Nov. 2020.

\bibitem{6G5}
S. Deng, H. Zhao, W. Fang, J. Yin, S. Dustdar, and A. Y. Zomaya, ``Edge intelligence: The confluence of edge computing and artificial intelligence," \emph{IEEE Internet of Things J.}, vol. 7, no. 8, pp. 7457-7469, Aug. 2020.

\bibitem{CL1}
W. Shi, J. Cao, Q. Zhang, Y. Li, and L. Xu, ``Edge computing: Vision and challenges," \emph{IEEE Internet Things J.}, vol. 3, no. 5, pp. 637-646, Oct. 2016.

\bibitem{CL2}
H. Lee, S. H. Lee, T. Q. S. Quek, and I. Lee, ``Deep learning framework for wireless systems: Applications to optical wireless communications," \emph{IEEE Commun. Mag.}, vol. 57, no. 3, pp. 35-41, Mar. 2019.

\bibitem{FEEL1}
G. Zhu, D. Liu, Y. Du, C. You, J. Zhang, and K. Huang, ``Towards an intelligent edge: Wireless communication meets machine learning," \emph{IEEE Commun. Mag.}, vol. 58, no. 1, pp. 19-25, Jan. 2020.


\bibitem{FL1}
B. McMahan, E. Moore, D. Ramage, S. Hampson, and B. A. Y. Arcas, ``Communication-efficient learning of deep networks from decentralized data," in \emph{Proc. Artif. Intell. Statist.}, 2017, pp. 1273-1282.

\bibitem{FL2}
T. Li, A. K. Sahu, A. Talwalkar, and V. Smith, ``Federated learning: Challenges, methods, and future directions," \emph{IEEE Signal Process.  Mag.}, vol. 37, no. 3, pp. 50-60, May 2020.

\bibitem{FL3}
B. McMahan, E. Moore, D. Ramage, S. Hampson, and B. A. Y. Arcas, ``Communication-efficient learning of deep networks from decentralized data," in \emph{Proc. Int. Conf. Artif. Intell. Stat. (AISTATS)}, 2017, pp. 1273-1282.

\bibitem{GBoard1}
A. Hard, K. Rao, R. Mathews, S. Ramaswamy et al, ``Federated Learning for Mobile Keyboard Prediction," [Online]. Available: https://arxiv.org/abs/1811.03604, 2018.

\bibitem{GBoard2}
S.Ramaswamy,  R. Mathews, K. Rao, and F. Beaufays, ``Federated learning for emoji prediction in a mobile keyboard," [Online]. Available: https://arxiv.org/abs/1906.04329, 2019.


\bibitem{FL4}
T. Li, A. K. Sahu, M. Zaheer, M. Sanjabi, A. Talwalkar, and V. Smith, ``Federated optimization in heterogeneous networks," in \emph{Proc. Machine Learning and Systems}, 2020, pp. 429-450.

\bibitem{FL5}
D. A. E. Acar, Y. Zhao, Navarro, R. M. Mattina, P. N. Whatmough, and V. Saligrama,``Federated learning based on dynamic regularization," [Online]. Available:  https://arxiv.org/abs/2111.04263, 2021.

\bibitem{ML0}
F. Sattler, S. Wiedemann, K.-R. M\"{u}ler, and W. Samek, ``Sparse binary compression: Towards distributed deep learning with minimal communication,"  in \emph{Proc. IEEE Int. Joint Conf. Neural Netw. (IJCNN)}, 2019, pp. 1-8.

\bibitem{ML1}
M. Chen, Z. Yang, W. Saad, C. Yin, H. V. Poor, and S. Cui, ``A joint learning and communications framework for federated learning over wireless networks," \emph{IEEE Trans. Wireless Commun.}, vol. 20, no. 1, pp. 269-283, Jan. 2021.

\bibitem{ML2}
N. Shlezinger, M. Chen, Y. C. Eldar, H. V. Poor, and S. Cui, ``UVeQFed: Universal vector quantization for federated learning," \emph{IEEE Trans. Signal Process.}, vol. 69, pp. 500-514, 2021.

\bibitem{OMA1}
J. Ren, Y. He, D. Wen, G. Yu, K. Huang, and D. Guo, ``Scheduling for cellular federated edge learning with importance and channel awareness," \emph{IEEE Trans. Wireless Commun.}, vol. 19, no. 11, pp. 7690-7703, Nov. 2020.

\bibitem{OMA2}
Q. Zeng, Y. Du, K. Huang, and K. K. Leung, ``Energy-efficient radio resource allocation for federated edge learning," in \emph{Proc. 2020 ICC Workshops}, Dublin, Ireland, 2020, pp. 1-6.

\bibitem{OMA3}
Z. Yang, M. Chen, W. Saad, C. S. Hong, and M. Shikh-Bahaei, ``Energy efficient federated learning over wireless communication networks," \emph{IEEE Trans. Wireless Commun.}, vol. 20, no. 3, pp. 1935-1949, Mar. 2021.

\bibitem{AirComp1}
O. Abari, H. Rahul, and D. Katabi, ``Over-the-air function computation in sensor networks,"  [Online]. Available: http://arxiv.org/abs/1612.02307, 2016.

\bibitem{AirComp2}
X. Chen and Q. Qi, \emph{Convergence of Energy, Computation and Communication in B5G Cellular Internet of Things}, Germany: Springer, 2020.

\bibitem{Nomofun}
M. Goldenbaum, H. Boche, and S. Staczak, ``Nomographic functions: Efficient computation in clustered gaussian sensor networks," \emph{IEEE Trans. Wireless Commun.}, vol. 14, no. 4, pp. 2093-2105, Apr. 2015.

\bibitem{Heterogeneous}
T. Sery, N. Shlezinger, K. Cohen, and Y. C. Eldar, ``Over-the-air federated learning from heterogeneous data," \emph{IEEE Trans. Signal Process.}, vol. 69, pp. 3796-3811, Jun. 2021.

\bibitem{Onebit}
G. Zhu, Y. Du, D. Gunduz, and K. Huang, ``One-bit over-the-air aggregation for communication-efficient federated edge learning: Design and convergence analysis," \emph{IEEE Trans. Wireless Commun.}, vol. 20, no. 3, pp. 2120-2135, Mar. 2021.

\bibitem{BBAFL}
G. Zhu, Y. Wang, and K. Huang, ``Broadband analog aggregation for low-latency federated edge learning," \emph{IEEE Trans. Wireless Commun.}, vol. 19, no. 1, pp. 491-506, Jan. 2020.

\bibitem{YKFL}
K. Yang, T. Jiang, Y. Shi, and Z. Ding, ``Federated learning via over-the-air computation," \emph{IEEE Trans. Wireless Commun.}, vol. 19, no. 3, pp. 2022-2035, Mar. 2020.

\bibitem{IRSFL}
Z. Wang, J. Qiu, Y. Zhou, Y. Shi, L. Fu, W. Chen, and K. B. Letaief, ``Federated learning via intelligent reflecting surface",  \emph{IEEE Trans. Wireless Commun.}, vol. 21, no. 2, pp. 808-822, Feb. 2022.

\bibitem{parallel}
M. M. Amiri, D. Guduz, S. R. Kulkarni, and H. Vincent Poor, ``Convergence of federated learning over a noisy downlink," \emph{IEEE Trans. Wireless Commun.}, vol. 21, no. 3, pp. 1422-1437, Mar. 2022.

\bibitem{Fading1}
X. Cao, G. Zhu, J. Xu, and K. Huang, ``Optimized power control for over-the-air computation in fading channels,"  \emph{IEEE Trans. Wireless Commun.}, vol. 19, no. 11, pp. 7498-7513, Nov. 2020.

\bibitem{CSI1}
H. Sun, F. Zhou, R. Q. Hu, and L. Hanzo, ``Robust beamforming design in a NOMA cognitive radio network relying on SWIPT," \emph{IEEE J. Sel. Areas Commun.}, vol. 37, no. 1, pp. 142-155, Jan. 2019.

\bibitem{CSI2}
Q. Qi, X. Chen, and D. W. K. Ng, ``Robust beamforming for NOMA-based cellular massive IoT with SWIPT," \emph{IEEE Trans. Signal Process.}, vol. 68, pp. 211-224, Dec. 2020.

\bibitem{Fading2}
M. M. Amiri and D. Gunduz, ``Federated learning over wireless fading channels," \emph{IEEE Trans. Wireless Commun.}, vol. 19, no. 5, pp. 3546-3557, May 2020.

\bibitem{AO}
J. C. Bezdek and R. J. Hathaway, ``Convergence of alternating optimization," \emph{Neural, Parallel \& Scientific Computations}, vol. 11, no. 4, pp. 351-368, Dec. 2003.


\bibitem{Convex}
S. Boyd and L. Vandenberghe, \emph{Convex Optimization}. Cambridge, U.K.: Cambridge Univ. Press, 2004.


\bibitem{CVX}
M. Grant and S. Boyd, \emph{CVX: Matlab Software for Disciplined Convex Programming.} [Online]. Available: http://cvxr.com/cvx.

\bibitem{AO1}
M. Razaviyayn, M. Hong, and Z. Q. Luo, ``A unified convergence analysis of block successive minimization methods for nonsmooth optimization," \emph{SIAM J. Optim.}, vol. 23, no. 2, pp. 1126-1153, Jun. 2013.

\bibitem{AO2}
J. W. Both, ``On the rate of convergence of alternating minimization for non-smooth non-strongly convex optimization in Banach spaces," [Online]. Available: https://arxiv.org/abs/1911.00404, Nov. 2019.


\bibitem{complexity}
A. Ben-Tal and A. Nemirovski, ``Lectures on modern convex optimization: Analysis, algorithms, and engineering applications,"  \emph{MPS-SIAMSeries on Optimization}. Philadelphia, PA, USA: SIAM, 2001.

\bibitem{NOMAbook}
X. Chen, \emph{Massive Access for Cellular Internet of Things: Theory and Technique}, Germany: Springer Press, 2019.

\bibitem{pathlossmodel}
\emph{Coordinated Multi-Point Operation for LTE Physical Layer Aspects (Rel. 11)}, document TR 36.819, 3GPP, Feb. 2011.

\bibitem{inequation}
G. H. Hardy, J. E. Littlewood, and G. Polya, \emph{Inequalities}.  Cambridge, U.K.: Cambridge Univ. Press, 1952.

\end{thebibliography}
\end{document}